\documentclass[lettersize,journal]{IEEEtran}
\usepackage{amsmath,amssymb,amsfonts}
\usepackage{algorithmic}
\usepackage{array}
\usepackage{textcomp}
\usepackage{caption}
\usepackage{subcaption}
\usepackage{stfloats}
\usepackage{verbatim}
\usepackage{graphicx}
\usepackage{balance}
\usepackage{xcolor}
\usepackage{physics}
\usepackage[hyphens]{url}
\usepackage{threeparttable}
\usepackage{booktabs}
\usepackage[nocompress]{cite}
\usepackage{bm}
\usepackage[acronyms, nonumberlist, nopostdot,nomain,nogroupskip]{glossaries}
\usepackage{amsmath}

\usepackage{tikz}
\usepackage{pgfplots}

\newacronym{3gpp}{3GPP}{3rd Generation Partnership Project}
\newacronym{adc}{ADC}{Analog to Digital Converter}
\newacronym{5g}{5G}{5th Generation}
\newacronym{6g}{6G}{6th Generation}
\newacronym{aimd}{AIMD}{Additive Increase Multiplicative Decrease}
\newacronym{am}{AM}{Acknowledged Mode}
\newacronym{amc}{AMC}{Adaptive Modulation and Coding}
\newacronym{aqm}{AQM}{Active Queue Management}
\newacronym{awgn}{AGWN}{Additive White Gaussian Noise}
\newacronym{balia}{BALIA}{Balanced Link Adaptation}
\newacronym{bdp}{BDP}{Bandwidth-Delay Product}
\newacronym{qos}{QoS}{Quality of Service}
\newacronym{qoe}{QoE}{Quality of Experience}
\newacronym{pqos}{PQoS}{Predictive Quality of Service}
\newacronym{bf}{BF}{Beamforming}
\newacronym{cc}{CC}{Congestion Control}
\newacronym{cu}{CU}{Centralized Unit}
\newacronym{ai}{AI}{Artificial Intelligence}
\newacronym{dql}{DQL}{Double Q-learning}
\newacronym{ppo}{PPO}{Proximal Policy Optimization}
\newacronym{du}{DU}{Distributed Unit}
\newacronym{cdf}{CDF}{Cumulative Distribution Function}
\newacronym{lidar}{LiDAR}{Light Detection and Ranging}
\newacronym{cn}{CN}{Core Network}
\newacronym{rl}{RL}{Reinforcement Learning}
\newacronym{cusum}{CUSUM}{CUmulative SUM}
\newacronym{cam}{CAM}{Cooperative Awareness Message}
\newacronym{mse}{MSE}{Mean Squared Error}
\newacronym{cqi}{CQI}{Channel Quality Information}
\newacronym[firstplural=Markov Decision Processes (MDPs)]{mdp}{MDP}{Markov Decision Process}
\newacronym{cp}{CP}{Control Plane}
\newacronym{csirs}{CSI-RS}{Channel State Information - Reference Signal}
\newacronym{dc}{DC}{Dual Connectivity}
\newacronym{dce}{DCE}{Direct Code Execution}
\newacronym{dci}{DCI}{Downlink Control Information}
\newacronym{dl}{DL}{Downlink}
\newacronym{dmr}{DMR}{Deadline Miss Ratio}
\newacronym{dmrs}{DMRS}{DeModulation Reference Signal}
\newacronym{e2e}{E2E}{end-to-end}
\newacronym{ecn}{ECN}{Explicit Congestion Notification}
\newacronym{edf}{EDF}{Earliest Deadline First}
\newacronym{enb}{eNB}{evolved Node Base}
\newacronym{epc}{EPC}{Evolved Packet Core}
\newacronym{es}{ES}{Edge Server}
\newacronym{fdma}{FDMA}{Frequency Division Multiple Access}
\newacronym{fdd}{FDD}{Frequency Division Duplexing}
\newacronym[firstplural=Radio Access Technologies (RATs)]{rat}{RAT}{Radio Access Technology}
\newacronym{fs}{FS}{Fast Switching}
\newacronym{ftp}{FTP}{File Transfer Protocol}
\newacronym{gnb}{gNB}{Next Generation Node Base}
\newacronym{harq}{HARQ}{Hybrid Automatic Repeat reQuest}
\newacronym{hetnet}{HetNet}{Heterogeneous Network}
\newacronym{hh}{HH}{Hard Handover}
\newacronym{hol}{HOL}{Head-of-Line}
\newacronym{ia}{IA}{Initial Access}
\newacronym{ieee}{IEEE}{Institute of Electrical and Electronics Engineers}
\newacronym{imt}{IMT}{International Mobile Telecommunication}
\newacronym{iot}{IoT}{Internet of Things}
\newacronym{ldpc}{LDPC}{Low-Density Parity Check}
\newacronym{los}{LOS}{Line-of-Sight}
\newacronym{lte}{LTE}{Long Term Evolution}
\newacronym{m2m}{M2M}{Machine to Machine}
\newacronym{nn}{NN}{Neural Network}
\newacronym{ml}{ML}{Machine Learning}
\newacronym{mac}{MAC}{Medium Access Control}
\newacronym{mc}{MC}{Multi-Connectivity}
\newacronym{mcs}{MCS}{Modulation and Coding Scheme}
\newacronym{mec}{MEC}{Mobile Edge Cloud}
\newacronym{mi}{MI}{Mutual Information}
\newacronym{mimo}{MIMO}{Multiple Input, Multiple Output}
\newacronym{mmwave}{mmWave}{millimeter wave}
\newacronym{mptcp}{MPTCP}{Multipath TCP}
\newacronym{mr}{MR}{Maximum Rate}
\newacronym{mss}{MSS}{Maximum Segment Size}
\newacronym{mtd}{MTD}{Machine-Type Device}
\newacronym{mtu}{MTU}{Maximum Transmission Unit}
\newacronym{nfv}{NFV}{Network Function Virtualization}
\newacronym{nlos}{NLOS}{Non-Line-of-Sight}
\newacronym{nlosv}{NLOSv}{Vehicle Non-Line-of-Sight}
\newacronym{nr}{NR}{New Radio}
\newacronym{ofdm}{OFDM}{Orthogonal Frequency Division Multiplexing}
\newacronym{pdcch}{PDCCH}{Physical Downlonk Control Channel}
\newacronym{pdcp}{PDCP}{Packet Data Convergence Protocol}
\newacronym{pdsch}{PDSCH}{Physical Downlink Shared Channel}
\newacronym{pdu}{PDU}{Packet Data Unit}
\newacronym{pf}{PF}{Proportional Fair}
\newacronym{ds}{D-S}{Dynamic Segmentation}
\newacronym{pgw}{PGW}{Packet Gateway}
\newacronym{phy}{PHY}{Physical}
\newacronym{pbch}{PBCH}{Physical Broadcast Channel}
\newacronym[plural=\gls{mme}s,firstplural=Mobility Management Entities (MMEs)]{mme}{MME}{Mobility Management Entity}
\newacronym{prb}{PRB}{Physical Resource Block}
\newacronym{pss}{PSS}{Primary Synchronization Signal}
\newacronym{pscch}{PSCCH}{Physical Sidelink Control Channel}
\newacronym{pucch}{PUCCH}{Physical Uplink Control Channel}
\newacronym{pusch}{PUSCH}{Physical Uplink Shared Channel}
\newacronym{rach}{RACH}{Random Access Channel}
\newacronym{ran}{RAN}{Radio Access Network}
\newacronym{red}{RED}{Random Early Detection}
\newacronym{rf}{RF}{Radio Frequency}
\newacronym{rlc}{RLC}{Radio Link Control}
\newacronym{rlf}{RLF}{Radio Link Failure}
\newacronym{rrc}{RRC}{Radio Resource Control}
\newacronym{rrm}{RRM}{Radio Resource Management}
\newacronym{rru}{RRU}{Remote Radio Unit}
\newacronym{rr}{RR}{Round Robin}
\newacronym{rs}{RS}{Remote Server}
\newacronym{rsrp}{RSRP}{Reference Signal Received Power}
\newacronym{rss}{RSS}{Received Signal Strength}
\newacronym{rtt}{RTT}{Round Trip Time}
\newacronym{rw}{RW}{Receive Window}
\newacronym{rx}{RX}{Receiver}
\newacronym{sa}{SA}{standalone}
\newacronym{sack}{SACK}{Selective Acknowledgment}
\newacronym{sap}{SAP}{Service Access Point}
\newacronym{sc}{SC}{Single Carrier}
\newacronym{sch}{SCH}{Secondary Cell Handover}
\newacronym{scoot}{SCOOT}{Split Cycle Offset Optimization Technique}
\newacronym{sdma}{SDMA}{Spatial Division Multiple Access}
\newacronym{sinr}{SINR}{Signal to Interference plus Noise Ratio}
\newacronym{sl}{SL}{Sidelink}
\newacronym{sm}{SM}{Saturation Mode}
\newacronym{snr}{SNR}{Signal-to-Noise-Ratio}
\newacronym{son}{SON}{Self-Organizing Network}
\newacronym{ss}{SS}{Synchronization Signal}
\newacronym{srs}{SRS}{Sounding Reference Signal}
\newacronym{sss}{SSS}{Secondary Synchronization Signal}
\newacronym{tb}{TB}{Transport Block}
\newacronym{tcp}{TCP}{Transmission Control Protocol}
\newacronym{tdd}{TDD}{Time Division Duplexing}
\newacronym{tdma}{TDMA}{Time Division Multiple Access}
\newacronym{tfl}{TfL}{Transport for London}
\newacronym{tm}{TM}{Transparent Mode}
\newacronym{trp}{TRP}{Transmitter Receiver Pair}
\newacronym{tti}{TTI}{Transmission Time Interval}
\newacronym{ttt}{TTT}{Time-to-Trigger}
\newacronym{tx}{TX}{Transmitter}
\newacronym{ue}{UE}{User Equipment}
\newacronym{ul}{UL}{Uplink}
\newacronym{uml}{UML}{Unified Modeling Language}
\newacronym{um}{UM}{Unacknowledged Mode}
\newacronym{utc}{UTC}{Urban Traffic Control}
\newacronym{vm}{VM}{Virtual Machine}
\newacronym{rsrq}{RSRQ}{Reference Signal Received Quality}
\newacronym{rssi}{RSSI}{Received Signal Strength Indicator}
\newacronym{crs}{CRS}{Cell Reference Signal}
\newacronym{nsa}{NSA}{Non Stand Alone}
\newacronym{mrdc}{MR-DC}{Multi \gls{rat} \gls{dc}}
\newacronym{endc}{EN-DC}{E-UTRAN-\gls{nr} \gls{dc}}
\newacronym{5gc}{5GC}{5G Core}
\newacronym{si}{SI}{Study Item}
\newacronym{iab}{IAB}{Integrated Access and Backhaul}
\newacronym{ibr}{IBR}{Iterative Best Response}
\newacronym{wf}{WF}{Wired-first}
\newacronym{hqf}{HQF}{Highest-quality-first}
\newacronym{pa}{PA}{Position-aware}
\newacronym{mlr}{MLR}{Maximum-local-rate}
\newacronym{wbf}{WBF}{Wired Bias Function}
\newacronym{mib}{MIB}{Master Information Block}
\newacronym{sib}{SIB}{Secondary Information Block}
\newacronym{rnti}{RNTI}{Radio Network Temporary Identifier}
\newacronym{dft}{DFT}{Discrete Fourier Transform}
\newacronym{kpi}{KPI}{Key Performance Indicator}
\newacronym{ppp}{PPP}{Poisson Point Process}
\newacronym{v2v}{V2V}{Vehicle-to-Vehicle}
\newacronym{wave}{WAVE}{Wireless Access in Vehicular Environments}
\newacronym{udp}{UDP}{User Datagram Protocol}
\newacronym{upa}{UPA}{Uniform Planar Array}
\newacronym{fec}{FEC}{Forward Error Correction}
\newacronym{v2x}{V2X}{Vehicle-To-Everything}
\newacronym{psfch}{PSFCH}{Physical Sidelink Feedback Channel}
\newacronym{pssch}{PSSCH}{Physical Sidelink Shared Channel}
\newacronym{csma}{CSMA}{Carrier Sense Multiple Access}
\newacronym{v2n}{V2N}{Vehicle-to-Network}
\newacronym{wlan}{WLAN}{Wireless Local Area Network}
\newacronym{cav}{CAV}{Connected and Autonomous Vehicle}
\newacronym{v2i}{V2I}{Vehicle-to-Infrastructure}
\newacronym{d2d}{D2D}{Device-to-Device}
\newacronym{c-its}{C-ITS}{Connected Intelligent Transportation System}
\newacronym{fr2}{FR2}{Frequency Range 2}
\newacronym{fr1}{FR1}{Frequency Range 1}
\newacronym{bs}{BS}{Base Station}
\newacronym{sdu}{SDU}{Service Data Unit}
\newacronym{csi}{CSI}{Channel State Information}
\newacronym{scs}{SCS}{Subcarrier Spacing}
\newacronym{sumo}{SUMO}{Simulation of Urban MObility}
\newacronym{prp}{PRP}{Packet Reception Probability}
\newacronym{dnn}{DNN}{Deep Neural Network}
\newacronym{pdr}{PDR}{Packet Delivery Ratio}
\newacronym{edca}{EDCA}{Enhanced Distribution Channel Access}
\newacronym{sdap}{SDAP}{Service Data Adaptation Protocol}
\newacronym{osm}{OSM}{OpenStreetMap}
\newacronym{rsu}{RSU}{Road Side Unit}
\newacronym{hv}{HV}{Host Vehicle}
\newacronym{imsi}{IMSI}{International Mobile Subscriber Identity}
\newacronym{lcid}{LCID}{Logical Channel Identifier}
\newacronym{nnet}{NN}{Neural Network}
\newacronym{movav}{MovAv}{Moving Average}
\newacronym{idw}{IDW}{Inverse Distance Weighting}
\newacronym{lin}{Lin}{Linear Interpolation}
\newacronym{relu}{ReLU}{Rectified Linear Unit}
\newacronym{adam}{Adam}{Adaptive moment estimator}
\newacronym{drl}{DRL}{Deep Reinforcement Learning}
\newacronym{fcd}{FCD}{Floating Car Data}
\newacronym{cd}{CD}{Chamfer Distance}
\newacronym{csv}{CSV}{Comma Separated Values}
\newacronym{ca}{CA}{Carrier Aggregation}
\newacronym{nwdaf}{NWDAF}{Network Data Analytics Function}
\newacronym{etsi}{ETSI}{European Telecommunications Standards Institute}
\newacronym{fl}{FL}{federated learning}
\newacronym{nef}{NEF}{Network Exposure Function}
\newacronym{5qi}{5QI}{5G QoS Identifier}
\newacronym{oam}{OAM}{Operations, Administration and Management}
\newacronym{sba}{SBA}{Service-based Architecture}
\newacronym{5gaa}{5GAA}{5G Automotive Association}
\newacronym{lan}{LAN}{Local Area Network}
\newacronym{rem}{REM}{Radio Environment Map}
\newacronym{lstm}{LSTM}{Long Short Term Memory}
\newacronym{svm}{SVM}{Support Vector Machine}
\newacronym{ns3}{ns-3}{Network Simulator 3}
\newacronym{gmv2}{GEMV$^2$}{Geometry-based Efficient propagation Model for V2V communication}
\newacronym{prada}{PRADA}{\underline{PR}edictive Quality of Service in the \underline{R}AN based on \underline{A}I for Teleoperated \underline{D}riving \underline{A}pplications}
\usepackage{hyperref}
\usepackage[capitalize]{cleveref}

\crefname{section}{Sec.}{Secs.}

\def\BibTeX{{\rm B\kern-.05em{\sc i\kern-.025em b}\kern-.08em
T\kern-.1667em\lower.7ex\hbox{E}\kern-.125emX}}

\newcommand{\edit}[1]{{\color{black} #1}}

\usepackage[font=small]{caption}
\usepackage[font=footnotesize]{subcaption}

\begin{document}

\title{PRATA: A Framework to Enable Predictive QoS in Vehicular Networks via Artificial Intelligence}
\author{Federico Mason, \textit{Member, IEEE}, Tommaso Zugno, Matteo Drago, Marco Giordani, \textit{Member, IEEE}, \\ Mate Boban, Michele Zorzi \textit{Fellow, IEEE}

\vspace{-0.5cm}

\thanks{Federico Mason, Matteo Drago, Marco Giordani, and Michele Zorzi are with the Department of Information Engineering (DEI) of the University of Padova, Italy. Email: \texttt{name.surname@unipd.it}. Tommaso Zugno and Mate Boban are with Huawei Technologies, Munich Research Center, Germany. Email: \texttt{name.surname@huawei.com}. This work was partially supported by the European Union under the Italian National Recovery and Resilience Plan (NRRP) Mission 4, Component 2, Investment 1.3, CUP C93C22005250001, partnership on “Telecommunications of the Future” (PE00000001 - program “RESTART”).~
}}

\maketitle

\begin{abstract}
\gls{pqos} makes it possible to anticipate QoS changes, e.g., in wireless networks, and trigger appropriate countermeasures to avoid performance degradation.
A promising tool for \gls{pqos} is given by \gls{rl}, a methodology that enables the design of decision-making strategies for stochastic optimization.
In this manuscript, we present PRATA, a new simulation framework to enable PRedictive QoS based on AI for Teleoperated driving Applications.
PRATA consists of a modular pipeline that includes (i) an end-to-end protocol stack to simulate the 5G \gls{ran}, (ii) a tool for generating automotive data, and (iii) an \gls{ai} unit to optimize \gls{pqos}  decisions.
To prove its utility, we use PRATA to design an \gls{rl} unit, named RAN-AI, to optimize the segmentation level of teleoperated driving data in the event of resource saturation or channel degradation.
Hence, we show that the RAN-AI entity efficiently balances the trade-off between QoS and \gls{qoe} that characterize teleoperated driving applications, almost doubling the system performance compared to baseline approaches. 
In addition, by varying the learning settings of the RAN-AI entity, we investigate the impact of the state space and the relative cost of acquiring network data that are necessary for the implementation of \gls{rl}.
\end{abstract}

\begin{IEEEkeywords}
Predictive Quality of Service (PQoS), Artificial Intelligence (AI), Reinforcement Learning (RL), vehicular networks, teleoperated driving.
\end{IEEEkeywords}

\section{Introduction}
\label{sec:introduction}

\begin{tikzpicture}[remember picture, overlay]
      \node[draw,minimum width=4in] at ([yshift=-1cm]current page.north)  {This manuscript has been accepted for publication in IEEE Transactions on Communications.};
\end{tikzpicture}

\glsresetall

\IEEEPARstart{T}{he} transition towards the next generation of wireless networks (6G) is making communications technologies more and more data-centric, data-dependent, and automated~\cite{giordani2020toward}.
In this context, the network will focus on the optimization of control services to promptly adapt to the users' needs in real time~\cite{letaief2019roadmap}.
This paradigm can be addressed via the introduction of \gls{ml} or other means of \gls{ai} inside the network, to enable communication services to autonomously reconfigure and optimize without human intervention.
Notably, \gls{ai} should not only address instant requirements, but also predict the evolution of the network to avoid \gls{qos} degradation and corresponding safety threats.

Over the past few years, \gls{ai} has been gaining increasing attention in vehicular networks, and \gls{ai}-assisted algorithms have been explored to support advanced applications like traffic and congestion control, platoon management, and autonomous driving~\cite{tong2019artificial}.
In particular, \gls{ai} allows to aggregate and process automotive data~\cite{giordani2019investigating} and take complex decisions, including object detection~\cite{rossi2021role}, and data compression~\cite{nardo2022point}.
To do so, vehicles need to acquire, process, and disseminate a huge amount of data generated by onboard sensors to achieve cooperative perception of the surrounding environment.
On the other hand, vehicular applications have very stringent \gls{qos} requirements, and are very demanding in terms of both channel capacity and end-to-end delay~\cite{zheng2015heterogeneous}.

In vehicular networks, the number of connected nodes, especially cars and pedestrians, and their positions and channel conditions, constantly evolve, making it challenging to maintain the \gls{qos} levels requested for \gls{v2x} communication~\cite{gyawali2020challenges}.
At the same time, not complying with \gls{qos} may lead to catastrophic events, such as a delay in detecting a neighboring vehicle or an error in performing a maneuver.
To face this challenge, it is fundamental to foresee network evolution and make proactive decisions, that are aware of future changes in the \gls{qos} levels~\cite{kousaridas2021qos}.
To this goal, the scientific community introduced the concept of \gls{pqos}, which aims at anticipating communication impairments and taking proper countermeasures to avoid service degradation~\cite{boban2021predictive}. 

For ensuring \gls{pqos} in vehicular networks, a promising tool is given by the \gls{rl} paradigm~\cite{kaelbling1996reinforcement}, which enables the design of complex decision-making strategies in stochastic environments. 
During the training phase, an \gls{rl} agent interacts with the environment according to a sequence of actions, each associated with an immediate reward.
Hence, the optimal strategy is identified by a trial and error process, that leads the agent to estimate the cumulative reward associated with each state-action combination~\cite{sutton2018reinforcement}.   

Notably, \gls{rl} has shown impressive results in several telecommunication scenarios~\cite{xiong2019deep}.
However, more than other learning approaches, \gls{rl} algorithms require the aggregation and processing of huge amounts of data to achieve convergence~\cite{osinski2020simulation}.  
In this regard, experiments with real testbeds are impractical due to limitations in scalability and flexibility, as well as the high cost of hardware components. 
The problem could be mitigated by collecting data from existing applications and performing the training according to an \emph{offline} approach~\cite{kumar2020conservative}. 
At the same time, the use of offline data may be inconvenient because of the nature of \gls{rl} agents, which are based on direct interactions with the environment. 

In this context, computer simulations are a fundamental tool for validating \gls{rl} algorithms or other \gls{ai} solutions and their interplay with communication networks.
Notably, most existing \gls{ai} frameworks are based on Python, which comes with many {ad hoc} libraries and solutions for training and optimization, like \texttt{PyTorch}.
However, most Python-based simulators rely on simplified assumptions that do not incorporate the complexity of real wireless communication systems. 
Instead, discrete-event network simulators, such as \gls{ns3}~\cite{henderson2008network}, have proven to be a solid alternative to analyze the performance of wireless networks without resorting to real testbeds.


In this manuscript, we present a novel simulation framework, named PRedictive QoS based on AI for Teleoperated driving Applications (PRATA), to design, evaluate, and dimension \gls{pqos} algorithms in vehicular networks.
The proposed system integrates and extends the \texttt{mmwave} module~\cite{mezzavilla2018end} of \gls{ns3} for the simulation of the channel and the 5G \gls{ran}, with an ad hoc Python tool to run and train \gls{ai} algorithms directly within the ns-3 code base.
This framework enables the optimization of PQoS solutions in a controllable environment based on the resulting impact on the \gls{ran}, and before the actual deployment on real vehicular applications.
In this sense, PRATA acts as a \textit{digital twin} of a vehicular network~\cite{zhao2022elite}, allowing network designers to investigate new protocols on a simulated, though accurate, environment during the training phase, with no risk of degrading the performance of the system.

To show the potential of PRATA, we analyze a teleoperated driving scenario, where vehicles exploit a 5G cellular connection to exchange data collected from onboard sensors with a remote driver.
Hence, we design a network orchestrator based on \gls{rl}, named RAN-AI, to select the communication setting(s) that ensure the optimal trade-off between \gls{qoe} and \gls{qos} requirements of the teleoperated driving application.
Specifically, the RAN-AI entity collects various communication metrics from the \gls{ran}, and the RL agent adjusts the segmentation mode, i.e., the size, of sensors' data accordingly.
Reducing the data size ensures a better \gls{qos} for the application, in terms of both delay and probability of reception, as data  can be transmitted faster, and the network is generally less congested.
However, adopting an excessively aggressive segmentation mode can jeopardize the quality of the received data, which may reduce the \gls{qoe} (measured, for example, in terms of the false alarm and/or misdetection probability of object detection on the data). 

\edit{Some of the results of this manuscript were presented at the \emph{2022 IEEE Wireless Communications and Networking Conference}~\cite{mason2022rlpqos}, and the \emph{2022 Workshop on ns-3}~\cite{drago2022wns3}. 
In particular, in~\cite{mason2022rlpqos}, we proposed for the first time the idea of using \gls{ai}/\gls{rl} to optimize \gls{pqos} operations for teleoperated driving, while in~\cite{drago2022wns3} we described our ns-3 module for the analyzing \gls{pqos} in vehicular networks.
This journal extends the contributions of our previous conference papers under several aspects, as we summarize below:}
\begin{itemize}   
    \item \edit{We formalize and improve the PQoS framework for teleoperated driving applications presented in~\cite{mason2022rlpqos} and~\cite{drago2022wns3}, which we now refer to as PRATA, focusing on the role of AI in the RAN to optimize PQoS operations.}
    
    \item \edit{We test PRATA in a large variety of scenarios and configurations, exploring the impact of the transmission power over the system performance, in addition to the network size, whose effect was already considered in our previous works.}
    
    \item  \edit{We compare the performance of RAN-AI against three static benchmarks (as done in~\cite{mason2022rlpqos} and~\cite{drago2022wns3}), where the segmentation level is pre-selected, and a novel heuristic algorithm which dynamically adjusts the segmentation mode of the teleoperated driving application according to the end-to-end delay.}
    \item \edit{We evaluate two training configurations for the RAN-AI entity, i.e., a \emph{value-based} approach based on Double Q-Learning (DQL), which was first described in~\cite{mason2022rlpqos}, and a new \emph{policy-based} approach based on Proximal Policy Optimization (PPO); we show that the latter choice is more robust for PQoS, especially as the number of vehicles increases.}
    \item \edit{Compared to our previous work, where the agent state was pre-defined, we investigate the benefits and drawbacks of using different state spaces, specifically the number and type of measurements available at the RAN-AI, with considerations related to the cost for acquiring and aggregating network data for the training, and the resulting learning~complexity.}
\end{itemize}
Our results demonstrate that PRATA, using AI, can improve PQoS performance in the teleoperated driving scenario in terms of both QoS and QoE, also in complex communication environments.
 In particular, PPO is a more convenient approach in multi-agent non-stationary systems, and outperforms DQL in terms of training performance.
At the same time, we demonstrate that increasing the size of the learning space may involve significant communication overhead, which can eventually degrade the performance of \gls{ai} protocols.

The rest of the paper is structured as follows.
In Sec.~\ref{sec:state_of_the_art} we review the state of the art on PQoS and AI in the context of vehicular and teleoperated driving networks.
In Sec.~\ref{sec:PRATA} we describe PRATA and its modules.
In Sec.~\ref{sec:ran-ai} we present the RAN-AI entity and describe its training setup.
In Sec.~\ref{sec:results} we evaluate the performance of PRATA in different teleoperated driving scenarios and conditions. 
Finally, in Sec.~\ref{sec:conclusions} we conclude the paper with suggestions for future~research.

\section{Related Work}
\label{sec:state_of_the_art}

The integration of \gls{ai} functionalities into \glspl{ran} is envisioned as a key enabler for the evolution of wireless systems towards 6G. Different papers highlighted the important role and identified the potential use cases of \gls{ai}-enabled \glspl{ran} \cite{khan2023ai, 9124820}, also focusing on vehicular communication scenarios \cite{10537699, rizzo2023towards}. One of the main applications is represented by \gls{pqos}, where \gls{ai} is used to predict the \gls{qos} performance of vehicular applications by measuring the \glspl{kpi} collected by the radio network. In this regard, several papers explored the potential and challenges of different \gls{ml}- and \gls{ai}-based solutions to accurately predict the performance of vehicular channels \cite{boban2021predictive}.

The authors of \cite{9860971} and \cite{jomrich2018cellular} evaluated different \gls{ml} approaches, including linear regression, deep \glspl{nnet}, and random forest, for predicting the throughput of connected vehicles. 
A UE-based \gls{qos} predictor was proposed in \cite{9129382} to forecast the E2E delay performance obtained by training a neural network model on real-world measurements. 
In \cite{9129097}, \gls{qos} predictions were used to derive the minimum inter-vehicle distance that should be kept to avoid safety risks, while in \cite{10012940}, the authors proposed a solution for proactive radio resource management operations based on \gls{pqos}. Different datasets for training and testing \gls{pqos} algorithms have been presented in \cite{10200750} and \cite{10268872}.

Among the many applications characterizing vehicular networks, teleoperated driving can gain great benefits from \gls{pqos} strategies.
In this regard, \cite{kousaridas2021qos} discussed the active standardization activities on \gls{pqos} and presented a \gls{qos} prediction scheme for teleoperated driving. 
In \cite{9604941}, a \gls{pqos} framework based on a \gls{lstm} model for uplink throughput predictions was presented, while  \cite{9685405} proposed an \gls{nnet} architecture for \gls{pqos} that incorporates prior knowledge about the communication system (e.g., cell load and channel condition) for teleoperated driving.
The authors of \cite{9773810} designed a framework for in-advance prediction of the end-to-end data rate of teleoperated vehicles based on radio environmental maps and presented a custom testbed for the evaluation of \gls{pqos} solutions.

Despite the great effort in the \gls{pqos} field, most research works still present issues that demand additional investigations. 
First, state-of-the-art solutions exploit pre-defined datasets and evaluate \gls{pqos} algorithms in an offline fashion.
Such an approach does not take into account the feedback effect that service adaptation has on \gls{qos} predictions that, therefore, may be inaccurate with respect to a real system~\cite{https://doi.org/10.48550/arxiv.2302.11966}.
In this work, PRATA addresses this problem by training the RAN-AI entity with sequences of data correlated in the time domain. 
This approach accurately mimics real scenarios where, after receiving \gls{pqos} notifications, the communication protocols are adapted to upcoming system changes, and the future \gls{qos} is re-estimated according to new information.

Besides, we highlight that most previous works focus on how to estimate future \gls{qos} requirements without considering how to adapt vehicular applications accordingly.
In this paper, we overcome this problem by adopting an integrated approach where prediction and adaptation are treated jointly.
PRATA enables both the collection of relevant \gls{pqos} metrics from the network environment and the reconfiguration of communication protocols to avoid performance degradation. 
Hence, the RAN-AI entity of PRATA exploits the collected metrics to estimate both the system state and the long-term performance associated with each communication decision. 
The final system makes it possible to analyze the mutual influence of predictions and countermeasures, enabling the definition of a comprehensive \gls{pqos} policy.

\section{Description of PRATA}
\label{sec:PRATA}

\begin{figure}[t!]
\vspace{-0.25cm}
\centering\includegraphics[width=0.75\columnwidth]{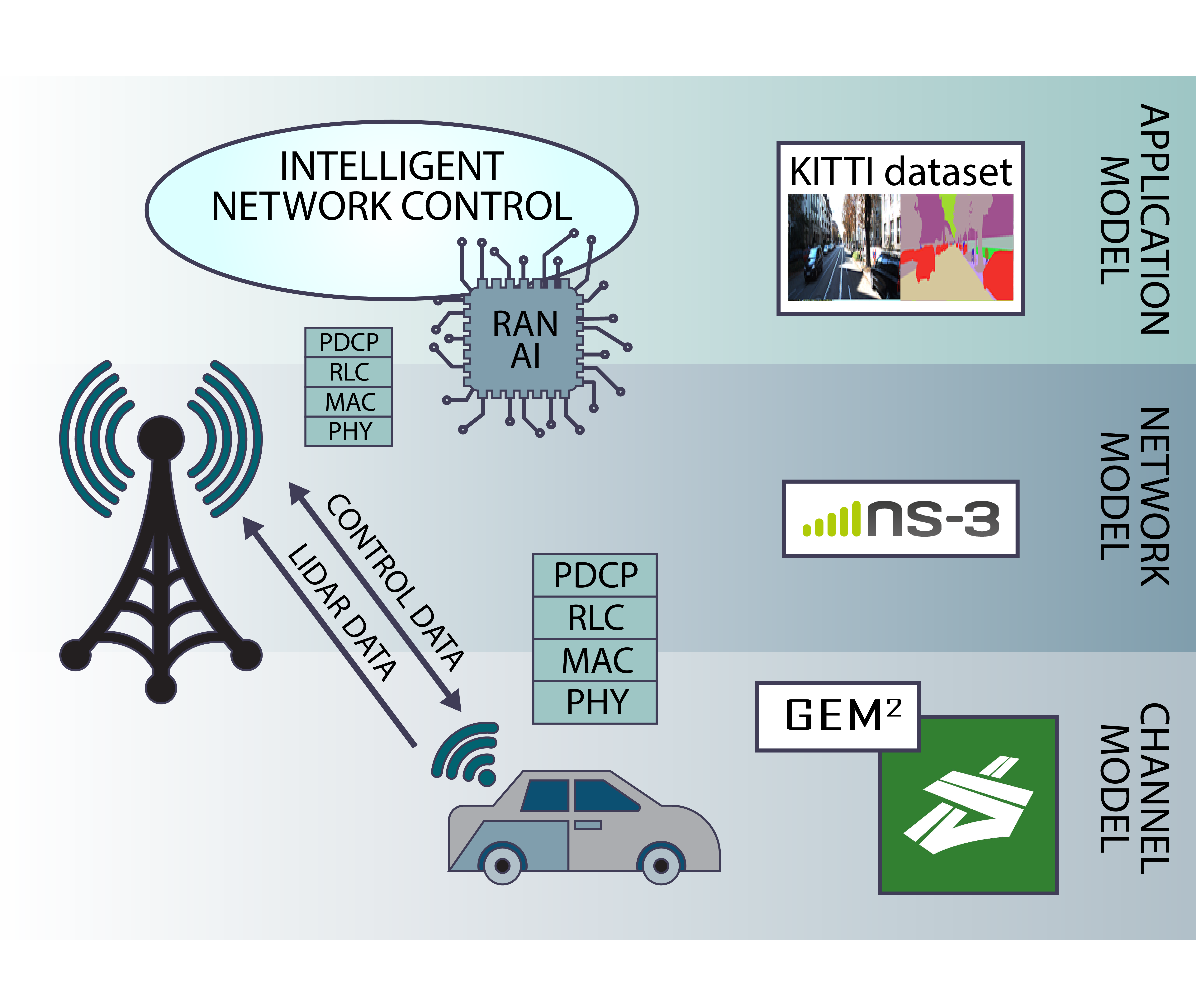}
\caption{Schematic overview of the PRATA framework.}
\label{fig:prata}
\vspace{-0.25cm}
\end{figure}

In this section, we describe the proposed PRATA framework for PQoS.\footnote{The source code of PRATA is publicly available at \texttt{https://github.com/signetlabdei/ns3-ran-ai}.}
PRATA is mainly based on \gls{ns3}, a well-known open-source software that models the different layers of the communication network's protocol stack.
To better characterize the  \gls{v2x} communication scenario, e.g., in terms of mobility and propagation, PRATA integrates \gls{ns3} with additional tools, including \gls{gmv2}~\cite{gemv2} and \gls{sumo}~\cite{SUMO2012}. 

In the rest of the section, we will describe the scenario where PRATA is tested, and the various modules of the framework, which is illustrated in Fig.~\ref{fig:prata}. 
Particularly, we can distinguish four building blocks: (i) the channel and mobility model, which accounts for the characterization of the wireless channel and the user mobility; (ii) the network model, which simulates the communication network; (iii) the application model, which simulates the teleoperated driving application; and (iv) the intelligent network controller, which provides \gls{ai} functionalities.
This latter block implements the RAN-AI entity, which will be better described in Sec.~\ref{sec:ran-ai}.

\subsection{Target Scenario}
\label{sub:scenario}

Although PRATA may be applied to many different use cases, we focus on a teleoperated driving system where multiple vehicles, i.e., the \glspl{ue}, are connected to the same \gls{gnb}.
In this scenario, an accurate \gls{pqos} strategy should guarantee that communication requirements, e.g., in terms of delay and packet loss, are not violated even in the case of channel quality degradation.
Hence, PRATA can be used to adapt the speed or trajectory of vehicles to steer them towards less crowded areas, and/or to offload network traffic to adjacent (less congested) cells~\cite{boban2021predictive}.
Another possible strategy may involve the optimization of some RAN parameters, especially at the application layer, to alleviate the burden of the channel before transmission.

In this work, we exploit PRATA to define a learning entity, named RAN-AI, that is installed at the \gls{gnb} and, thus, constitutes an integral element of the \gls{ran}. 
The RAN-AI entity is connected to the different \gls{ran} components, from the \gls{cu} to the \glspl{rru}, as well as to the \gls{cn}, by means of dedicated interfaces.
We emphasize that the RAN-AI entity acts across different layers of the protocol stack, from the physical layer to the application layer, so the effects of its decisions may impact various elements of the network in a holistic fashion.

Specifically, we assign to the RAN-AI entity the role of optimizing the segmentation mode, i.e., the size, of the data acquired by the \gls{lidar} sensors of the teleoperated vehicles under the coverage of the target \gls{gnb}.
In particular, the data segmentation software assigns labels to \gls{lidar} acquisitions, and removes redundant and/or irrelevant data points, thereby reducing the size of the data at the application layer before transmission.
As such, the RAN-AI entity operates at the application layer, and orchestrates the data exchange between the vehicles and the cellular network.
We emphasize that this optimization is fundamental in the context of this work, where network capacity is limited, which requires the application to reduce the data rate in the presence of network congestion.

We recall that PRATA is \gls{ai}-agnostic, meaning that different \gls{ai} algorithms can be seamlessly implemented into the simulation framework for the optimization of the system.
In this work, the RAN-AI entity is designed according to a centralized architecture, so that training and inference are both performed at the \gls{gnb}. 
However, the design of PRATA can be extended to consider distributed and/or \gls{fl} solutions, where the learning models are trained directly onboard the teleoperated vehicles, as occurs in an Edge-\gls{ai} scenario~\cite{bragato2023towards,bragato2024federated}. 
In particular, this approach prevents raw data to be exchanged between vehicles and the \gls{gnb} during the inference phase, which promotes faster processing and privacy.
On the other hand, centralized learning generally leads to more accurate results since data is aggregated from multiple users, and centralized servers are generally equipped with more powerful computing units for the training.

\subsection{Channel and Mobility Model}
\label{sub:channel}

To accurately characterize vehicular communication, it is fundamental to obtain a realistic representation of both the road network and the vehicles' trajectories.  
To reach the first goal, PRATA exploits \gls{osm}, a free and openly available database of geographical data, enabling the modeling of the road map of real locations. 
Then, the \gls{osm} representation of the scenario is processed by \gls{sumo}, an open-source simulator that models vehicular scenarios according to a discrete time frame. 
In particular, \gls{sumo} provides ad hoc functions for acquiring \gls{osm} data and generating realistic routes of vehicles moving in the target scenario.

To compute the wireless channel, PRATA processes the information from the environment and vehicles' trajectories through \gls{gmv2}, an open-source geometry-based propagation model for \gls{v2x} scenarios~\cite{boban2014geometry}.
Practically, \gls{gmv2} calculates the channel fading components on a large and small scale, and outputs the propagation loss for all possible vehicle pairs at each time step, taking into account their speed and relative position with respect to \gls{gnb}. 
The channel traces are then fed to \gls{ns3} through an ad hoc script reading the output of \gls{gmv2} and computing the power received between any pair of devices in the target scenario.
The overall process is implemented through the \texttt{propagation} module of \gls{ns3}.


\subsection{Network Model}
\label{sub:network}

To emulate communication across different layers of the protocol stack, PRATA uses and integrates \texttt{mmwave}, an {\gls{ns3}} module that has been taken as a reference simulator for 5G scenarios~\cite{mezzavilla2018end}.
Although \texttt{mmwave} was originally designed to simulate data communication at \gls{mmwave} frequencies, i.e., in \gls{fr2}, in PRATA we use \texttt{mmwave} to also operate in the lower part of the spectrum, i.e., in \gls{fr1}, as also supported in the 5G NR standard. 
In particular, PRATA integrates the \texttt{mmwave} module with \gls{gmv2}, making it possible to implement \glspl{gnb} and \glspl{ue} without beamforming capabilities.

Within the PRATA framework, the \gls{phy} and \gls{mac} layers are customized to support multiple NR-compliant frame structures, beamforming algorithms, and scheduling policies.
Instead, the \gls{rlc} and \gls{pdcp} layers are extended to include the functions of the \texttt{lena} module for \gls{lte} networks~\cite{piro2011lte}. 
Finally, PRATA supports dual connectivity with \gls{lte} base stations, which enables the simulation of non-standalone 5G deployments, and \gls{ca} at the \gls{mac} layer. 



\subsection{Application Model}
\label{sub:application}

To emulate teleoperated driving applications, PRATA features a traffic model which generates data packets as a function of the size of the frames acquired by the sensors of the vehicles, the segmentation and/or compression strategy applied to the frames, and the time periodicity at which frames are exchanged.
In PRATA, the size of frames is modeled based on {Kitti}, a popular dataset of vehicular perception data, collected using a Volkswagen Passat equipped with a Velodyne \gls{lidar}, two RGB cameras, and two grey-scale cameras~\cite{geiger2012are}. 
In this work, we focus exclusively on \gls{lidar} sensors, as they are generally more challenging to handle, and expose more interesting research questions.

For processing \gls{lidar} data, PRATA exploits the same pipeline proposed in~\cite{varischio2021hybrid} and, specifically, implements a semantic segmentation of point clouds with RangeNet++~\cite{rnet}, considering the following segmentation modes.
\begin{itemize}
    \item \emph{Raw (R)}: no data points are removed, and the raw LiDAR acquisition is considered.
    \item \emph{Segmentation Conservative (SC)}: data points associated with road elements are removed.  
    \item \emph{Segmentation Aggressive (SA)}: data points associated with road elements, buildings, vegetation and the background are removed, thus keeping only the most critical features (typically pedestrians and vehicles).
\end{itemize}

To simulate data transmission, PRATA extends the \texttt{TraceFileBurstGenerator} tool presented in \cite{lecci21bursty}.
In particular, PRATA associates each data frame with an ad hoc attribute that indicates the size of the frame and the time interval between the transmission of the current frame and the following one, i.e., the inter-frame rate. 
Within the \texttt{mmwave} module, two ad hoc functions take care of the fragmentation of data intro bursts, and re-aggregation. 

Finally, PRATA models and incorporates the processing time for segmentation and compression of data frames before transmission to the end users. 
This feature is also implemented on the receiver side to model the decoding time for processing the packets that reach the destination. 
Particularly, the delay associated with each segmentation operation is determined according to the results obtained in our previous work~\cite{varischio2021hybrid}, studying point cloud compression.

\subsection{Intelligent Network Controller} 
\label{sub:intelligent_controller}

As explained at the beginning of this section, PRATA enables the optimization of the \gls{ran} via an ad hoc entity, named RAN-AI, which acts as a middleware between the network and the teleoperated vehicles.
This entity provides an efficient data exchange between Python and \gls{ns3}, exploiting a shared memory for inter-process communications and offering a high-level interface in both Python and C++. 

Practically, the RAN-AI entity is implemented via an ad hoc \gls{ns3} module, named \texttt{RanAI}, that enables the optimization (based on AI) of the teleoperated vehicles associated with a specific \gls{gnb}.
This module continuously executes a periodic routine to allow the \gls{gnb} to aggregate the measurements used by the RAN-AI entity, and make network decisions (i.e., the segmentation mode to use) accordingly.
As we will better explain in Sec.~\ref{sec:ran-ai}, the aggregation of the communication measurements referred to a single vehicle represents, for the RAN-AI entity, the state of the vehicle itself. 
To enable the collection of such measurements, PRATA extends the ns-3 \texttt{mmwave} with new attributes that accommodate interactions with the \texttt{RanAI} module.

The RAN-AI entity operations are not directly managed by \gls{ns3}, but by a custom Python module, named \texttt{CentralizedAgent}.
This module interacts with \texttt{RanAI} and periodically receives all the measurements collected in the \gls{gnb}. 
From a practical perspective, \texttt{CentralizedAgent} implements two main functions:
\begin{itemize}
  \item The first takes as an input the measurements of all the vehicles connected with the target \gls{gnb}, and returns a decision for each one of them;
  \item The second makes the RAN-AI entity perform a new training step, refining its policy according to the information collected up to that moment. 
\end{itemize}
We observe that the proposed implementation is agnostic to the network size, and could be implemented for any number of teleoperated vehicles. 
Consequently, PRATA can be exploited to evaluate the network performance in scenarios characterized by extreme conditions, in terms of connectivity requests and resource consumption.


\section{The RAN-AI Entity in PRATA}
\label{sec:ran-ai}
The RAN-AI entity in PRATA implements an \gls{rl} agent, whose goal is to identify the segmentation mode that maximizes the \gls{qoe} of each remote driver without violating the \gls{qos} requirements of the network.
Practically, the agent is encouraged to select more aggressive segmentation to reduce the data size, and so the data rate of the application, in case of bad channel conditions, if the impact in terms of QoE degradation is within certain bounds. 
In the following subsections, we will introduce the reward function for the training and evaluation of the system.
Then, we will recall the main \gls{rl} fundamentals, and explain the methodological details behind the \gls{rl} agent's operations in the  RAN-AI.  

\subsection{Problem and Performance Definition}
\label{sub:reward}

In teleoperated driving scenarios, performance analysis is not straightforward, due to the complexity of vehicle interactions and of the channel model. 
In our framework, the performance of the \emph{target vehicle} coincides with the reward $r(\cdot)$ of the \gls{rl} agent managing such a vehicle, so the RAN-AI entity aims at maximizing the vehicle's reward over the long term. 
Therefore, in the rest of the paper the terms reward and performance are used interchangeably.
In particular, the performance is modeled in terms of both \gls{qos}, which depends on the communication requirements, and \gls{qoe}, which depends on the quality of the data received by the end user after segmentation
\footnote{QoE is formally defined in~\cite{perkis2020qualinet} as the degree of delight or annoyance of the user of an application or service, i.e., it has to do with the subjective user experience.
In this study, to highlight the trade-off between the quality of the data (as per the compression decisions made by the application) and the network performance (which depends, among other things, on the channel quality and the network traffic), we adopt a slightly different definition, where QoE is the quality of the data after segmentation, which is intuitively related to the user experience if the network had perfect performance.
By this approach, we can directly study the trade-off between data compression and network congestion, both of which ultimately affect the user's experience.}.

\edit{The \gls{qos} depends on the end-to-end delay $\delta$ and the \gls{prp} associated with the transmission process.
We define the $\text{\gls{prp}}$ of the target vehicle at time $t$ as the ratio between the number of packets received by the \gls{gnb}, and the total number of packets transmitted by the vehicle:
\begin{equation}
    {\text{\gls{prp}}}(t) = \frac{N_{\text{rx}}(t)}{N_{\text{tx}}(t)}
\label{eq:qos}
\end{equation}
Hence, the \gls{qos} at time $t$ is $1$ if the delay ${\delta}(t)$ and packet reception probability ${\text{\gls{prp}}}(t)$ associated with the target vehicle comply with the \glspl{kpi} for teleoperated driving applications, and zero otherwise, i.e.,}
\begin{equation}
	\begin{aligned}
	\mathrm{QoS}(t) =
        \begin{cases}
        1 \quad & \delta(t) \leq \delta_M \, \wedge \, \text{\gls{prp}}(t) \geq \text{\gls{prp}}_m, \\
        0 \quad & \text{otherwise},  
	\end{cases}
	\label{eq:qos}
	\end{aligned}
\end{equation}
\edit{where $\delta_M$ and $\text{\gls{prp}}_m$ are the maximum tolerable end-to-end delay and the minimum tolerable \gls{prp}, as specified in~\cite{3GPP_22186}\footnote{\edit{Notice that, in the 3GPP specifications~\cite{3GPP_22186}, the term ``packet reception probability'' is referred to as ``reliability.''}.}}

The \gls{qoe} denotes whether the quality of \gls{lidar} data is good enough for teleoperated driving applications, e.g., object detection~\cite{rossi2021role}.
To measure quality, we consider the symmetric point-to-point \gls{cd}~\cite{varischio2021hybrid} between the transmitted $\hat{D}$ and the original \gls{lidar} data $D$:
\begin{equation}
	\begin{aligned}
	\mathrm{CD} = \sum_{\forall \mathbf{d}\in D} \min_{\forall \hat{\mathbf{d}}\in\hat{D}} \lVert \mathbf{\mathbf{d}} - \hat{\mathbf{d}} \rVert_2^2 + \sum_{\forall \hat{\mathbf{d}}\in \hat{D}} \min_{\forall \mathbf{d}\in D} \lVert \mathbf{ \mathbf{d}} - \hat{\mathbf{d}} \rVert_2^2.
	\label{eq:chamfer}
	\end{aligned}
\end{equation}
The value of the \gls{qoe} at time $t$ gets lower as the actual Chamfer Distance $\mathrm{CD}(t)$ increases:
\begin{equation}
	\begin{aligned}
	\mathrm{QoE}(t) = \frac{\mathrm{CD}_{M} -\mathrm{CD}(t)}{\mathrm{CD}_{M}},
	\label{eq:qoe}
	\end{aligned}
\end{equation}
where $\mathrm{CD}_{M}$ is the maximum \gls{cd} that can be supported by the user application.

\edit{Notably, our \gls{qoe} model is still approximate as the minimization of the \gls{cd} does not ensure that the user application is necessarily accurate.
More realistic \gls{qoe} definitions may evaluate the quality of the application based on the ability of an end user to detect critical road elements from the received data, which may be still possible even if the \gls{cd} takes large values.
However, the study of more advanced methodologies to define and compute \gls{qoe} in the context of teleoperated driving is out of the scope of this manuscript and will be be left for future research.
We refer the interested readers to~\cite{brunnstrom2013qualinet} to further explore this subject.}

To make our system capture both QoS and QoE metrics, we define a piece-wise reward function, which returns $0$ whenever the \gls{qos} requirements are not satisfied, or a positive value that depends on both \gls{qos} and \gls{qoe} otherwise, i.e.,
\begin{equation}
\label{eq:reward}
r(t) =
\begin{cases}
(1-\alpha) \frac{\delta_M - \delta(t)}{\delta_M} + \alpha \mathrm{QoE}(t) \quad & \mathrm{QoS}(t)=1; \\
0 \quad & \mathrm{QoS}(t)<1.    
\end{cases} 
\end{equation}
In the above equation, $\alpha$ is a tuning parameter in $[0,1]$ that allows us to prioritize \gls{qos} over \gls{qoe} or \emph{vice-versa}; in any case, the performance (and, equivalently, the agent reward) always returns an output in $[0,1]$. 

\subsection{Reinforcement Learning}
\label{sub:learn_theory}

The RAN-AI is trained according to the \gls{rl} paradigm.
It is a powerful mathematical framework that models a scenario as a \gls{mdp} $(\mathcal{S}, \mathcal{A}, P(\cdot), R(\cdot))$, where $\mathcal{S}$ is the state space, $\mathcal{A}$ is the action space, $P: \mathcal{S} \times \mathcal{S} \times \mathcal{A} \rightarrow [0,1]$ is the state transition probability, and $R: \mathcal{S} \times \mathcal{S} \times \mathcal{A} \rightarrow \mathbb{R} $ is the reward function.
In particular, $P(s, s', a)$ is the probability that the system state evolves from $s$ to $s'$ because of action $a$, while $R(s, s', a)$ is the reward received by the agent after such state transition. 
Hence, we discretize time into slots and, in each slot~$t$, the agent observes the system state $s(t) \in \mathcal{S}$, takes a new action $a(t) \in \mathcal{S}$, and receives a reward $r(t)=R(s(t), s(t+1), a(t)) \in \mathbb{R}$ that depends on the new state $s(t+1) \in \mathcal{S}$ of the system.

The goal of any \gls{rl} agent is to determine the optimal policy $\pi^*$, i.e., the state-action pair probability $\pi: \mathcal{S} \times \mathcal{A} \rightarrow [0,1]$ that maximizes the mean cumulative reward
\begin{equation}
    G(t) = \sum_{\tau=t}^{\infty} \lambda^{\tau-t} r(\tau),
\end{equation}
where $r(t)$ is the reward received in slot $t$ and $\lambda \in (0,1)$ is the so-called \emph{discount factor}. 
To reach this goal, the agent defines a function $V_\pi: \mathcal{S} \rightarrow \mathbb{R}$, named \emph{value-function}, to estimate the quality of each possible state $s \in \mathcal{S}$.
Specifically, $V_\pi(s)$ represents the estimate of the cumulative return $G(t)$ that is obtained following the learned policy $\pi$ from state $s \in \mathcal{S}$, i.e., $V_\pi(s) = \mathbb{E}\left[ G(t) | s(t)=s, \pi \right]$.
The agent may also define a \emph{Q-function} $Q_\pi: \mathcal{S} \times \mathcal{A} \rightarrow \mathbb{R}$, which estimates the quality of the single state-action pair $(s, a) \in \mathcal{S} \times \mathcal{A}$, so that $Q_\pi(s,a) = \mathbb{E}\left[ G(t) | s(t)=s, a(t)=a, \pi \right]$.
The \emph{Bellman equations} ensure that the optimal policy $\pi^*$ leads to the highest value for both $V_\pi(s)$ and $Q_\pi(s, a)$ in each possible state and action.

As done in our previous work~\cite{mason2022rlpqos}, the RAN-AI entity can be trained according to the \gls{dql} algorithm described in~\cite{van2016deep}, which is an extended version of the classical \emph{Q-learning} with improved performance.
Even though \emph{value-based} methods are not recent, \gls{dql} still represents a valid tool for optimizing \gls{rl} scenarios with a discrete action space, as the one considered in this paper.
On the other hand, \gls{dql} may struggle to discover the optimal policy for non-stationary environments, which is the case when multiple vehicles are deployed~\cite{zakharenkov2021deep}.
For this reason, we also implement the \gls{ppo} algorithm~\cite{pmlr-v48-mniha16}, one of the most popular policy-gradient methods, thus obtaining two versions of the learning system with possibly different outcomes.
A detailed comparison between \gls{dql} and \gls{ppo} will be provided in Sec.~\ref{sec:results}.

\subsection{State and Action Space} 
\label{sub:state_and_action}

In our scenario, we assume that the RAN-AI models the application of each vehicle as a separate \gls{mdp} with identical statistics, and aggregates the data from all the vehicles to learn the optimal policy.
In other words, the same policy $\pi(\cdot)$ is applied to each vehicle connected to the \gls{gnb}, but each vehicle is associated with a separate state $s \in \mathcal{S}$.
By doing so, we implicitly assume that the RAN-AI's decisions referred to a specific vehicle do not influence the stochastic properties, i.e., the transition probability $P(\cdot)$ and the reward function $r(\cdot)$, of the \glspl{mdp} associated with the other vehicles.

The above assumption makes the state and action spaces independent of the network size, which permits to exploit the same learning architecture for any number of teleoperated vehicles.
We can take advantage of the RAN-AI scalability to investigate extreme scenarios, verifying how network architectures react when resource consumption is pushed to the limit. 
However, this design choice may increase the training variance since, by definition, the segmentation modes change during the training itself, which means that state observations obtained at different training times may be associated with different rewards.
To avoid possible performance degradation, it is therefore fundamental to accurately describe the state space of each vehicle.

Indeed, the RAN-AI entity can compute the vehicle states considering the following data:
\begin{itemize}
\item the \emph{context information}, which includes (i) the overall driving scenario, (ii) the network deployment, (iii) the vehicle trajectories (if available), and (iv) the time of the day or, possibly, the weather conditions;
\item the \emph{network metrics}, which include information about the physical, \gls{rlc} and \gls{pdcp} layers of the protocol stack, such as \gls{snr}, \glspl{prb} utilization, and \gls{mcs} index;
\item the \emph{application metrics}, including the \gls{e2e} mean, standard deviation, minimum and maximum values of delay and throughput.
\end{itemize}
By increasing the number of input metrics, we can enhance the RAN-AI's understanding of the environment, reducing the noise of the training process.
However, this choice increases the dimension of the state space $\mathcal{S}$ and, consequently, the complexity of the learning architecture. 
In Sec.~\ref{sec:results} we will vary the agent's input information, and evaluate the trade-off between the learning accuracy and the cost of the training.

We highlight that increasing the state space may not always improve the accuracy of the training phase, and a more robust approach would be to train a single vehicle at a time while leaving the other policies frozen, as done by \gls{ibr} strategies \cite{mason2024multi}. 
However, even \gls{ibr} promotes no convergence guarantees in our scenario, which consists of a general Markov game where finding a single Nash equilibrium is an NP-hard problem.
Another approach to improve the training process would be to implement feature compression and selection techniques, which allow the agents to reduce the size of the state space.
However, this solution is out of the scope of this work, and will be left for future research.

For what concerns the agent decisions, the action space $\mathcal{A}$ includes the segmentation modes $\{\rm R, SC, SA\}$ introduced in Sec.~\ref{sub:application}, by which the teleoperated vehicles process \gls{lidar} data before data transmission~\cite{nardo2022point}.
Hence, the RAN-AI entity has to determine which segmentation mode is more appropriate for ensuring that \gls{qos} requirements are not violated while maximizing \gls{qoe}.
Each action $a \in \{\rm R, SC, SA\}$ is thus associated with a different trade-off between \gls{qos} and \gls{qoe}, which depends on the current state of the environment. 
It is important to highlight that PRATA does not prevent other countermeasures from being considered, granting the interested researcher full flexibility.

\subsection{Learning Architecture}
\label{sub:learn_model}

To implement the RAN-AI, we consider a \gls{drl} approach, which means that the agent policy is approximated by an \gls{nnet}.
By doing so, we make it possible to handle continuous state spaces and overcome the \emph{curse of dimensionality} phenomenon~\cite{sutton2018reinforcement}.
For \gls{dql}, the agent is implemented by an \gls{nnet} with $S$ inputs and $A$ outputs.
Hence, the \gls{nnet} input size ($S$) is the number of input metrics of the RAN-AI entity, while the output size ($A$) is the number of agent's actions.
Instead, for \gls{ppo}, the agent is associated with two \glspl{nnet}, namely the \emph{critic} and the \emph{actor}.
The latter presents the same architecture used by \gls{dql}, while the critic presents $S$ inputs and $1$ output neuron, respectively.

All the learning units are implemented as feed-forward \glspl{nnet} with two hidden layers of $64$ and $16$ neurons, respectively, considering the \gls{relu} as a non-linear activation. 
This choice, together with the low dimensionality of the \gls{nnet}, ensures that the RAN-AI requires minimal computational capacity to be implemented. 
When \gls{dql} is used, the learning agent performs a total of $S \times 64 + 64 \times 16 + 16 \times 3$ operations to process each state observation, which generally involves a negligible effort for modern \glspl{gnb}.
However, the number of computational steps scales linearly with $S$ and the number of vehicles, which increases the learning complexity in case of congested networks.

At each step $t$, the RAN-AI receives the state $s(t) \in \mathcal{S}$ of each vehicle, and computes the Q-value $Q_\pi(s(t), a)$ associated with each action $a \in \mathcal{A}$ in the case of \gls{dql}, or directly the best action to take in the case of \gls{ppo}.
At the beginning of the training, the RAN-AI will mainly perform random actions, making it possible to explore the learning environment.
Once the exploration phase has ended, the RAN-AI will converge toward the action leading to the highest cumulative reward.
We recall that the sizes of the \gls{nnet} vary according to the state and action spaces, leading to a trade-off between the accuracy of the learned policy and the training complexity. 

\begin{table*}[t!]
\centering
\footnotesize
\caption{Simulation and learning parameters.}
\label{tab:params}
\begin{tabular}{lll|lll}
  \toprule
  Parameter & Description & Value & Parameter & Description & Value \\
  \midrule
  $f_c$ & Carrier frequency & 3.5 GHz & $B$ & Total bandwidth & 50 MHz \\ 
  $P_{tx}$ & Transmission power & $\{23, 30\}$ dBm & $T_{\text{update}}$ & RAN-AI update periodicity & 100 ms \\
  $T_{\text{episode}}$ & Episode duration & 80 s & $N_u$ & Number of vehicles & $\{1,\,5,\,10\}$ \\
  $\lambda$ & Discount factor  & 0.95 & $\zeta$ & DQL learning rate & $10^{-4}$ \\
  $N_{\text{epoch}}$ & PPO epochs per episode  & 32 & $\xi$ & PPO learning rates (actor and critic) & $\{1, 5\} \times 10^{-4}$ \\
  $\kappa$ & PPO entropy coefficient  & 0.01 & $\upsilon$ & PPO clip value & 0.2 \\ 
  $\{\Delta_{\text{up}}, \Delta_{\text{low}}\}$ & CUSUM strategy thresholds & $\{ 62.5, 37.5\}$ ms & $\alpha$ & Reward function weight & 1.0 \\
  $\delta_M$ & Max. tolerated delay & $50$ ms & \gls{prp}$_m$ & Min. tolerated \gls{prp} & $1.0$ \\
  $\mathrm{CD}_{M}$ & Max. tolerated \gls{cd}  & 45 & NN$_{\text{size}}$ & NN size & $S \times 64 \times 16 \times A $ \\
  $S$ & State space dimension & $\{5, 8, 10, 16, 18\}$ & $A$ & Action space size & $3$ \\
  $N_{\text{train}}$ & Training episodes & $10000 / N_u$ & $N_{\text{test}}$ & Testing episodes & $500 / N_u$ \\  
  \bottomrule
  \end{tabular}
  \vspace{-0.2cm}
\end{table*}

To overcome the computational effort associated with training the \gls{nnet}, and with the asynchronous exchange between \gls{ns3} and Python, PRATA enables the possibility of training the RAN-AI entity \textit{offline}.
In this case, the RAN-AI's actions are fixed for the whole training, while the learned policy is updated at the end of the simulation.  
However, as discussed in Sec.~\ref{sec:state_of_the_art}, purely offline training often does not converge to an optimal policy since it does not ensure a comprehensive exploration of the state-action spaces.
In the case of offline training, it is convenient to perform a batch of \emph{online} episodes, where actions follow some exploration policies, e.g., $\epsilon$-greedy, before implementing the system in a real scenario.

\section{Numerical Results}
\label{sec:results}

In this section, we exploit PRATA to analyze the teleoperated driving scenario introduced in Sec.~\ref{sec:PRATA}. 
Simulation and learning parameters are reported in Tab.~\ref{tab:params}.
\smallskip

\emph{Benchmarks.}
We compare the performance of PRATA using the RAN-AI entity (trained according to either the \gls{dql} or the \gls{ppo} algorithm) for the optimization of the segmentation mode against three constant benchmarks that select a static segmentation mode at the beginning of the simulation:
\begin{itemize}
    \item Constant Raw (CR), which transmits the raw data (segmentation mode: R).
    \item Constant Segmentation Conservative (C-SC), which transmits raw data without the road elements (segmentation mode: SC).
    \item Constant Segmentation Aggressive (C-SA), which only transmits critical data points (segmentation mode:~SA).
\end{itemize}
We also implement an additional heuristic benchmark, named \gls{ds}, which monitors the end-to-end delay of the vehicles via a \gls{cusum} control chart~\cite{chang1995cumulative}.
Whenever the cumulative delay is above an upper threshold $\Delta_{\text{up}}$, the \gls{ds} strategy turns toward a more aggressive segmentation mode.
If the cumulative delay drops below a lower threshold $\Delta_{\text{low}}$, a more conservative segmentation mode is selected.
By doing so, \gls{ds} tries to maximize the \gls{qoe} while not excessively degrading the \gls{qos}, similarly to what is done by the RAN-AI entity.
The performance of \gls{ds} depends on the hyperparameters $\Delta_{\text{up}}$ and~$\Delta_{\text{low}}$.
\smallskip

\emph{Learning convergence.}
To estimate the number of episodes necessary to ensure the agent's convergence, we considered, as primary indicators, the hierarchy between the Q-values (for \gls{dql}) and the entropy of the agent's policy (for \gls{ppo}). 
We empirically observed that these metrics stabilize when the RAN-AI collects around $5000 \times N_{\text{step}}$ state transitions from the environment, where $N_{\text{step}}$ is the number of steps per episode. 
Hence, we tune the training duration to collect a total of $10^4 \times N_{\text{step}}$ transitions, while the learning rate is deliberately small to prevent instability.
Notably, this approach still does not ensure that the final policy is optimal as the convergence of any \gls{rl} algorithm using \glspl{nnet} as policy approximators cannot be proved analytically.
\smallskip

\emph{Learning configuration.}
For each system configuration, we train the RAN-AI entity for $N_{\text{train}}= 10^4 / N_u$ episodes, where $N_u$ is the number of teleoperated vehicles.
By doing so, we take advantage of the fact that the learning process is faster when more users share the same experience.
Following the same principle, during the test phase, we carry out a total of $N_{\text{test}}=500 / N_u$ episodes for each configuration.
In the following, we always refer to the aggregate results obtained during the test phase, analyzing the statistical distributions of specific performance metrics.
\smallskip

\emph{Application.}
For the teleoperated driving application described in Sec.~\ref{sub:application}, LiDAR sensors generate point clouds at 10 fps, while the frame size depends on the segmentation mode.
For R, the size is 200 KB, for SC it is 100 KB, for SA it is 18 KB~\cite{varischio2021hybrid}.
The resulting data rates are 2, 1 and 0.18 MBps, respectively.
\smallskip

\emph{Performance metrics and parameters.}
The system performance is evaluated in terms of: (i) the mean reward as defined in Eq.~\eqref{eq:reward}; (ii) the mean QoS, defined as the probability to satisfy QoS requirements as in Eq.~\eqref{eq:qos}; (iii) the mean QoE as defined in Eq.~\eqref{eq:qoe}; (iv) the end-to-end delay; and (v) the \gls{prp}.
Moreover, we consider the following performance parameters: (i) the number $N_u$ of teleoperated vehicles; (ii) the transmit power $P_{{tx}}$; and (iii) the state space $\mathcal{S}$ at the RAN-AI.

\subsection{Impact of the Network Size}
\label{sub:size}

\begin{figure}[t!]
\centering
    \centering
    \includegraphics[width=.89\linewidth]{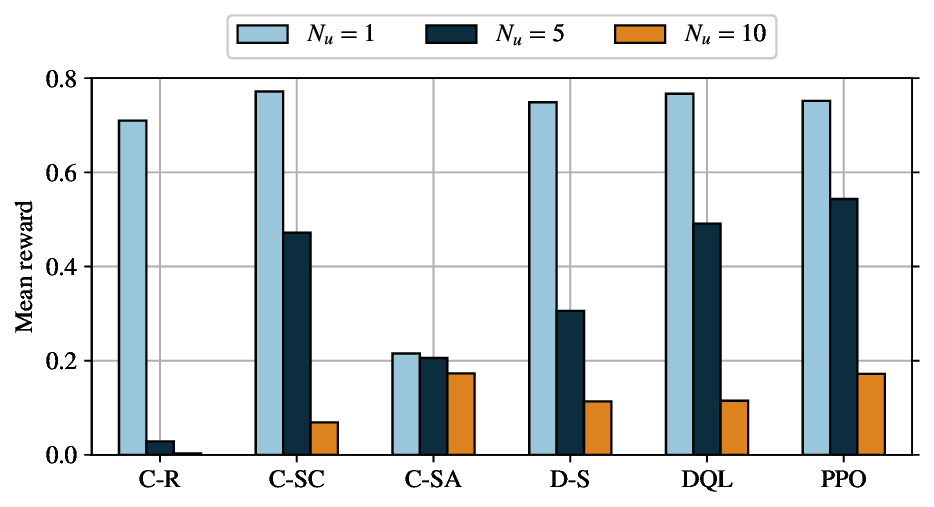}
    \caption{Mean reward as a function of $N_u$.}
    \label{fig:user_reward_bar}
    \vspace{-0.2cm}
\end{figure}

\begin{figure}[t!]
\centering
    \centering
    \includegraphics[width=.89\linewidth]{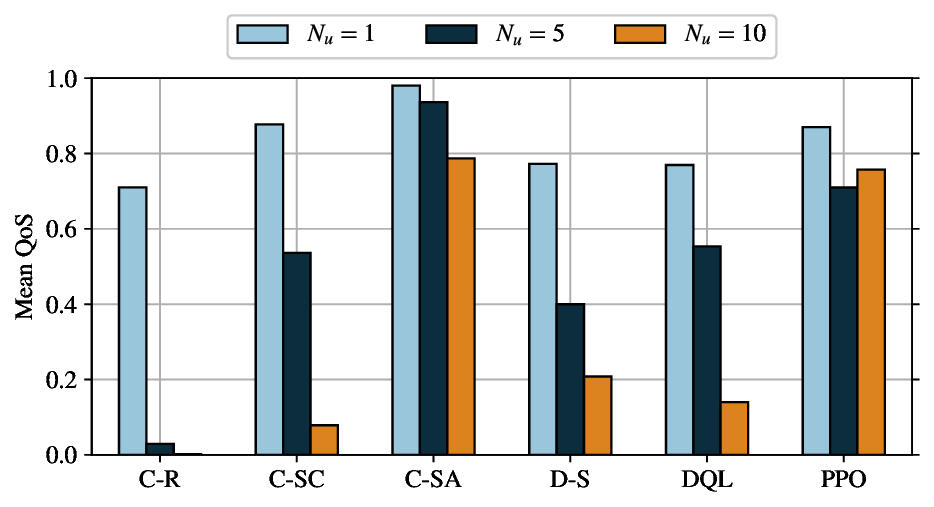}
    \caption{Mean QoS as a function of $N_u$..}
    \label{fig:user_qos_bar}
    \vspace{-0.2cm}
\end{figure}

In this initial set of results, we assume that the RAN-AI exploits all the input measurements generated at any level of the protocol stack (see Sec.~\ref{sub:state_and_action}), so the state of each vehicle consists of $S=18$ variables.
We fix the transmission power to $P_{tx}=30$ dBm, and evaluate the system performance versus the number of teleoperated vehicles $N_u \in \{1, 5, 10\}$.
Hence, in Fig.~\ref{fig:user_reward_bar}, we report the mean reward obtained during the testing phase using PRATA and the RAN-AI entity, considering both DQL and PPO as training algorithms, and the benchmark schemes.
We recall that the reward coincides with the system performance since it depends on both \gls{qos} and \gls{qoe}.

The results show that transmitting only raw data (C-R) ensures good performance when only one user is active, but the reward drops to almost zero when $N_u>1$.
To address more congested scenarios with multiple users, it is necessary to reduce the data rate via SC or SA segmentation. 
In doing so, the C-SC approach consistently improves the mean reward for $N_u=1$ and 5, but still shows poor performance when $N_u=10$.
Conversely, C-SA adopts the most aggressive segmentation, so the reward is almost twice that of C-SC for $N_u=10$.
However, the performance of C-SA deteriorates for $N_u=1$ and $5$ given the negative effect of segmentation on the QoE component of the reward.

Unlike the constant benchmarks, our heuristic (\gls{ds}) and RAN-AI-based solutions (DQL and PPO) try to adapt the segmentation mode to the actual network conditions.
In particular, \gls{ppo} overcomes the constant benchmarks in all cases, except when  $N_u=1$, where C-SC has slightly better performance.
When comparing the two training algorithms, we notice that \gls{ppo} outperforms \gls{dql} as $N_u$ increases.
We can explain this behavior by observing that \gls{dql} is a \emph{value-based} method, and may not handle well non-stationary environments, such as those characterized by multiple vehicles. 
On the other hand, \gls{ppo} represents a \emph{policy-based} method, and is known for providing higher performance and sample efficiency in multi-agent games than more conventional \gls{rl} algorithms~\cite{yu2022surprising}.
In any case, both DQL and PPO show superior performance than \gls{ds}, which demonstrates the potential of AI in this scenario.

\begin{figure}[t!]
    \centering
    \begin{subfigure}{.89\linewidth}
        \centering
        \includegraphics[width=.99\linewidth]{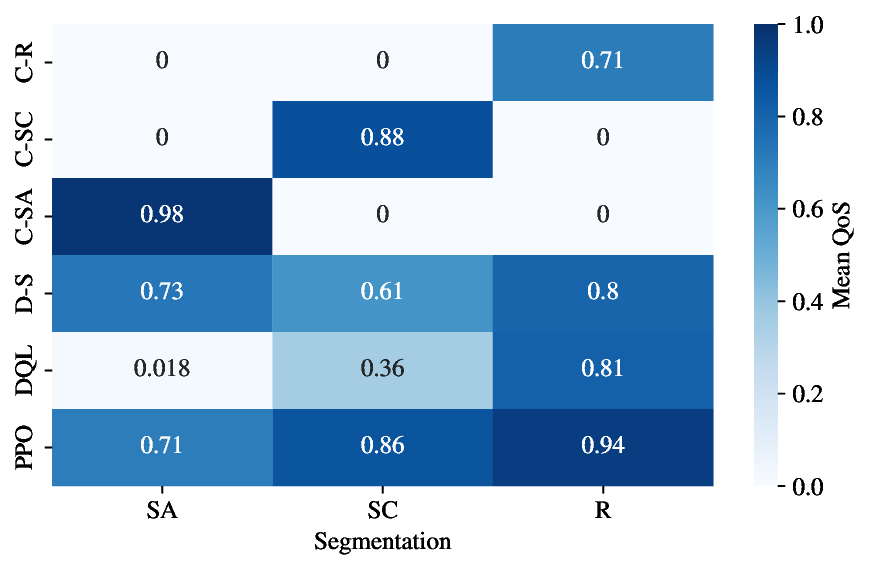}
        \caption{$N_u=1$.}
        \label{fig:user_qos_vs_qoe_1}
    \end{subfigure}
    \begin{subfigure}{.89\linewidth}
        \centering
        \includegraphics[width=.99\linewidth]{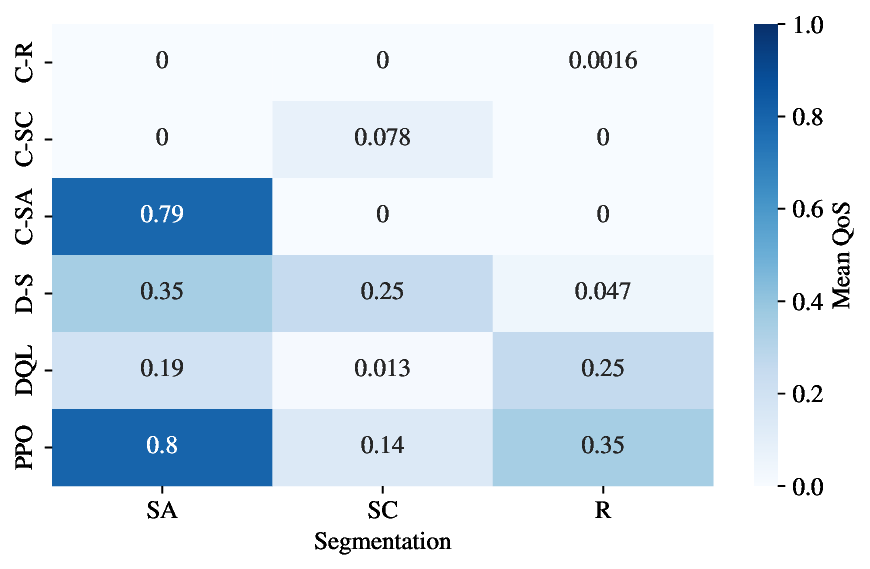}
        \caption{$N_u=10$.}
        \label{fig:user_qos_vs_qoe_10}
    \end{subfigure}
    \caption{Relation between QoS and QoE as a function of $N_u$.}
    \label{fig:user_qos_vs_qoe}
    \vspace{-0.2cm}
\end{figure}

Fig.~\ref{fig:user_qos_bar} shows that the mean \gls{qos} generally decreases as the network size increases, and more users contend for the same resources. 
Interestingly, \gls{qos} and reward are almost identical when using the C-R approach: this is because this technique does not apply any segmentation, and QoS is penalized (in favor of \gls{qoe}), especially as $N_u$ increases.
On the opposite side, the C-SA approach tends to maximize \gls{qos} via segmentation at the expense of QoE, independently of the number of vehicles, and it is as high as $0.79$ even when $N_u=10$.
For C-SC, D-S, and \gls{dql}, the mean QoS is greater than $0.75$ when $N_u=1$ but falls below $0.2$ when $N_u=10$.
On the other hand, \gls{ppo} radically improves the performance compared to \gls{dql}, even in the case of scenarios with multiple vehicles, and the mean \gls{qos} is still $0.76$ when $N_u=10$.

To provide a better overview of the relation between \gls{qos} and \gls{qoe}, Fig.~\ref{fig:user_qos_vs_qoe} investigates the mean \gls{qos} as a function of both the number of vehicles and the segmentation mode.
We recall that, by definition, D-S, \gls{dql}, and \gls{ppo} are designed to dynamically optimize the segmentation mode, while constant benchmarks select the segmentation mode at the beginning of the simulation.
As previously observed, the mean \gls{qos} generally decreases when $N_u$ increases, and the C-SA strategy ensures the best \gls{qos} independently of the number of vehicles.
We appreciate that \gls{ppo} leads to a higher \gls{qos} than all the competitors except when $N_u=1$, where C-SC and C-SA show slightly better results.
Indeed, \gls{ppo} has learned to reduce the level of segmentation, and so to improve the resulting QoE, only when channel conditions are good enough to address the system's \glspl{kpi} in terms of QoS; the same occurs for D-S and DQL which, however, have a lower \gls{qos}, on average.

\begin{figure}[t!]
    \centering
    \begin{subfigure}{0.89\linewidth}
        \centering
        \includegraphics[width=.99\linewidth]{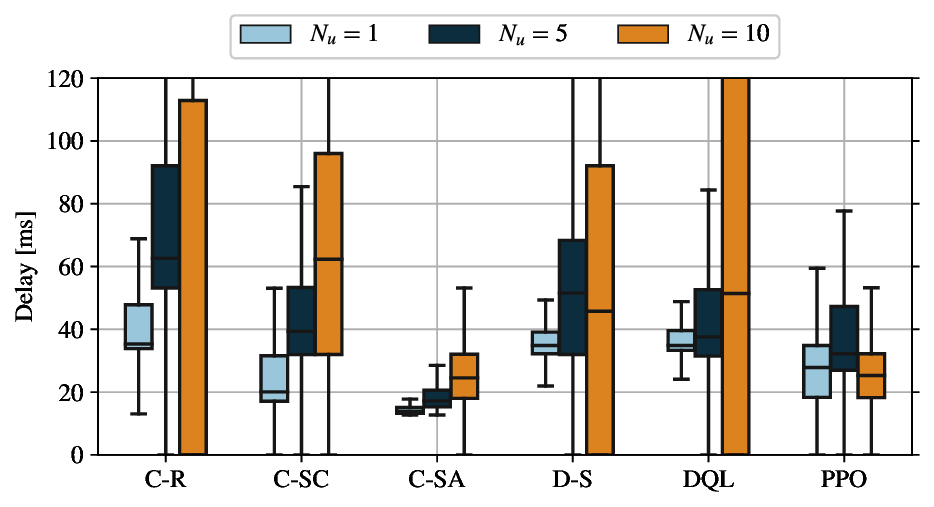}
        \caption{Delay distribution.}
        \label{fig:user_delay_box}
    \end{subfigure}\\
    \begin{subfigure}{0.89\linewidth}
        \centering
        \includegraphics[width=.99\linewidth]{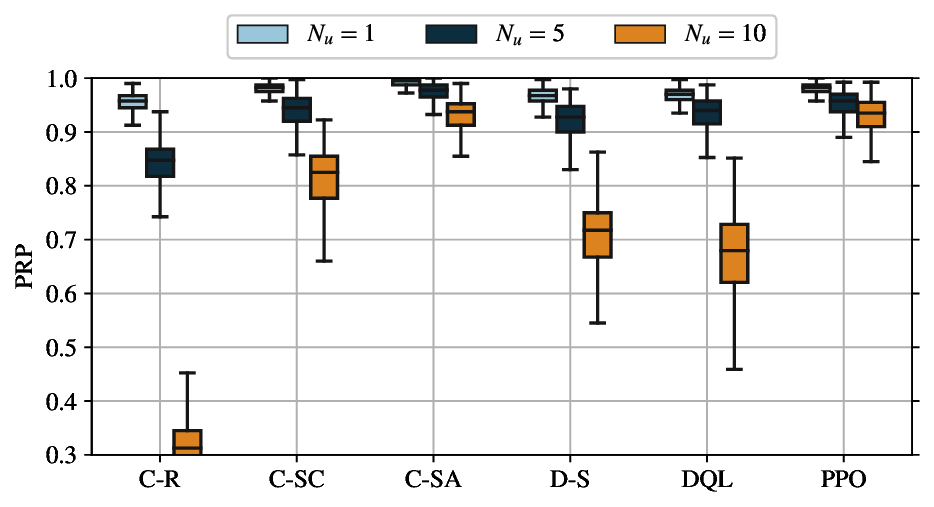}
        \caption{PRP distribution.}
        \label{fig:user_reception_bar}
    \end{subfigure}
    \caption{Communication metrics as a function of $N_u$.}
    \label{fig:user_communication_metric}
    \vspace{-0.2cm}
\end{figure}

Finally, in Fig.~\ref{fig:user_communication_metric}, we report the distribution of the end-to-end delay and \gls{prp}, measured at the application layer, under the same network configurations. 
To this goal, we use the boxplot representation, where the black line in the middle of each box represents the distribution median and the box edges are the $25$th and $75$th percentiles, while the box whiskers denote the outliers. 
As expected, the delay increases as more vehicles are in the system, while the \gls{prp} decreases under the same conditions. 
Particularly, the C-SA strategy delivers most of the packets in less than $40$ ms, independently of the number of vehicles, which is below the $\delta_M=50$ ms threshold imposed by the \glspl{kpi}.
In contrast, C-R fails to meet the delay requirements when $N_u>1$, and C-SC seems to provide acceptable results only when $N_u \leq 5$.
The \gls{dql} solution offer a balance between the different options, leading to a median delay lower than $40$ ms when $N_u \in \{1, 5\}$, and lower than $55$ ms when $N_u=10$.
Interestingly, \gls{ppo} has a similar delay than C-SA, but at the same time offers higher performance in terms of \gls{qoe} (as shown in Fig.~\ref{fig:user_qos_vs_qoe}).

In terms of \gls{prp}, C-SA and \gls{ppo} ensure that most packets are delivered in all scenarios, with a median \gls{prp} higher than $0.90$.
Performing no segmentation is extremely unfavorable as $N_u$ increases: the C-R strategy delivers less than $85\%$ of packets when $N_u=5$ and less than $35\%$ when $N_u=10$.
Finally, C-SC, D-S, and \gls{dql} have similar performance, delivering less than $85\%$ of packets when $N_u=10$, and more than $90\%$ in the other conditions. 

\subsection{Impact of the Transmission Power}
\label{sub:power}

\begin{figure}[t!]
\centering
    \centering
    \includegraphics[width=.89\linewidth]{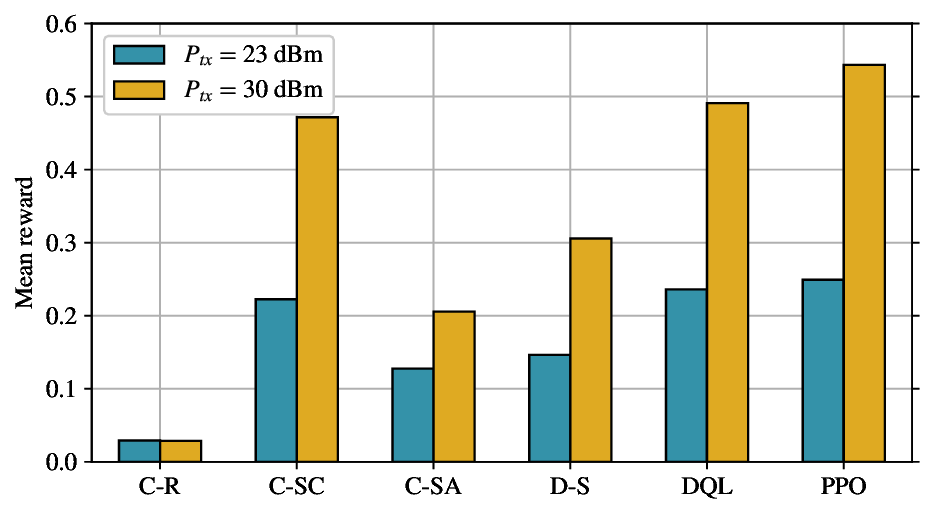}
    \caption{Mean reward as a function of $P_{tx}$.}
    \label{fig:power_reward_bar}
    \vspace{-0.2cm}
\end{figure}

Next, we analyze the performance of the RAN-AI fixing the number of users to $N_u=5$, and varying the transmission power $P_{tx} \in \{23, 30\}$ dBm.
As shown in Fig.~\ref{fig:power_reward_bar}, the channel quality is worse when $P_{tx}=23$ dBm compared to $P_{tx}=30$ dBm, so it is more challenging to satisfy to communication \glspl{kpi}, which leads to a lower mean reward.
Using the  RAN-AI with DQL and PPO, PRATA outperforms all the constant benchmarks independently of the power level, denoting the positive impact of \gls{ai} for optimizing PQoS. In fact, PRATA and the RAN-AI allow to adaptively select the segmentation mode based on the value of $P_{tx}$, compared to the constant benchmarks that apply segmentation regardless of the actual channel conditions.
Using the heuristic method for adapting the segmentation mode is not sufficient: for \gls{ppo} and \gls{dql}, the mean reward is more than twice that of \gls{ds} when $P_{tx}=30$ dBm, and one third higher when $P_{tx}=23$ dBm.
In particular, \gls{ppo} outperforms \gls{dql}, which further demonstrates our previous observation: in multi-user systems, policy-based methods like PPO are more robust.

\begin{figure}[t!]
    \centering
    \begin{subfigure}{0.89\linewidth}
        \centering
        \includegraphics[width=.99\linewidth]{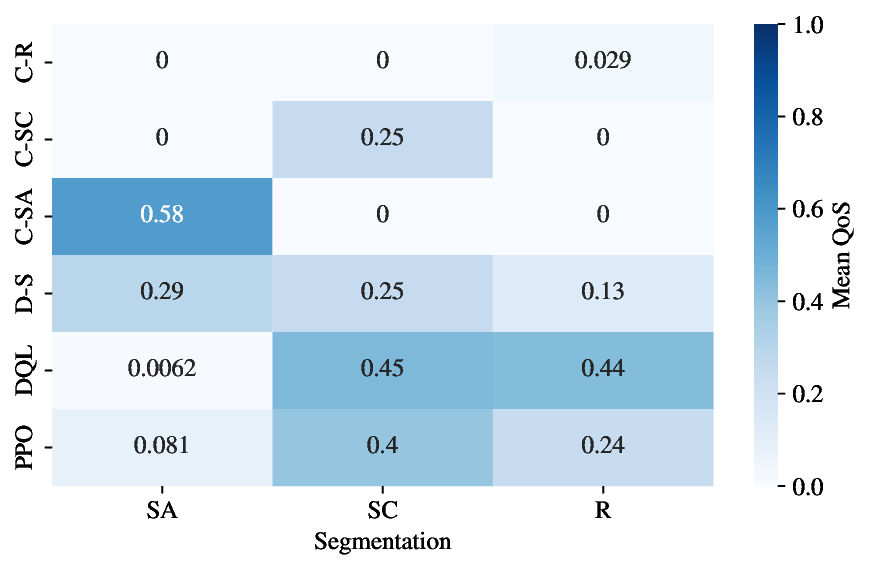}
        \caption{$P_{tx}=23$ dBm.}
        \label{fig:power_qoe_vs_qos_23}
    \end{subfigure}
    \\
    \begin{subfigure}{0.89\linewidth}
        \centering
        \includegraphics[width=.99\linewidth]{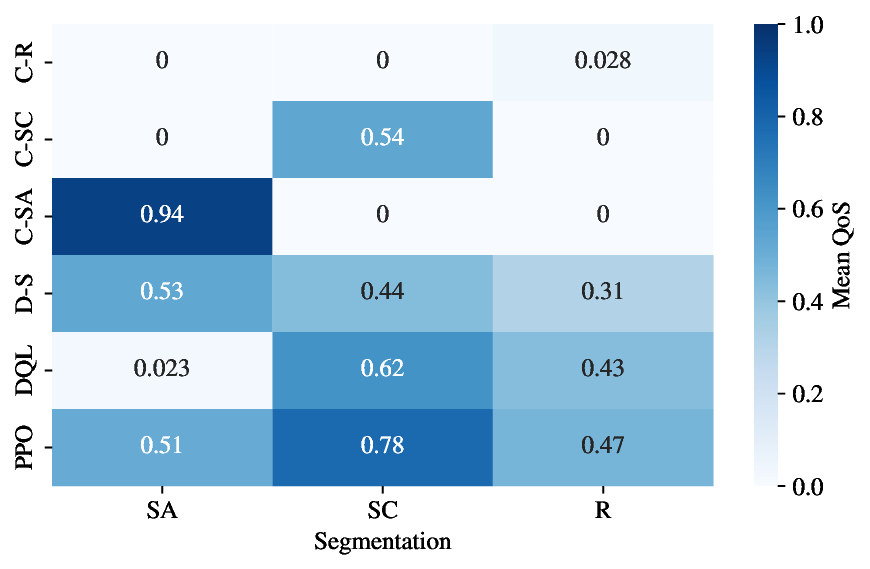}
        \caption{$P_{tx}=30$ dBm.}
        \label{fig:power_qoe_vs_qos_30}
    \end{subfigure}
    \caption{Relation between QoS and QoE as a function of $P_{tx}$.}
    \label{fig:power_qos_vs_qoe}
    \vspace{-0.2cm}
\end{figure}

In Fig.~\ref{fig:power_qos_vs_qoe} we study the relation between \gls{qos} and the segmentation mode as a function of $P_{tx}$.
When $P_{tx}=30$ dBm, \gls{ppo} outperforms all the other strategies in all conditions, except for C-SA which gives a mean \gls{qos} of $0.94$ despite a severe performance degradation in terms of QoE. This is confirmed by Fig.~\ref{fig:power_reward_bar}, which shows how PPO doubles the reward obtained with C-SA.
When $P_{tx}=23$ dBm, \gls{dql} leads to a mean \gls{qos} of $0.40$ for both the SC and R segmentation modes, whereas none of the benchmark strategies can ensure a similar performance.
On the other hand, \gls{ppo} strongly improves the mean \gls{qos} compared to \gls{dql} when the SA mode is adopted.  
Hence, \gls{ppo} is more conservative, and enables the management of more critical network conditions, as in the case of a limited transmission power.

\begin{figure}[t!]
    \centering
    \begin{subfigure}{0.89\linewidth}
        \centering
        \includegraphics[width=.99\linewidth]{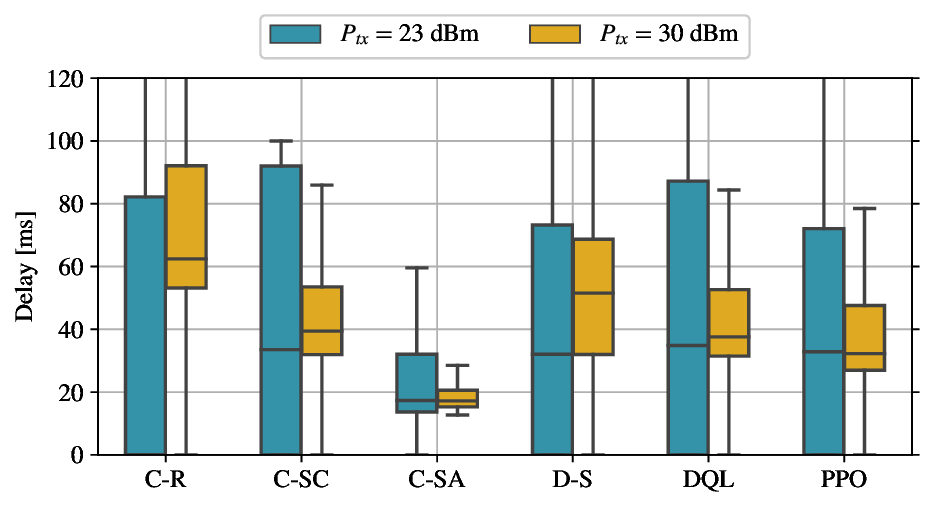}
        \caption{Packet delay distribution.}
        \label{fig:power_delay_box}
    \end{subfigure}\\
    \begin{subfigure}{0.89\linewidth}
        \centering
        \includegraphics[width=.99\linewidth]{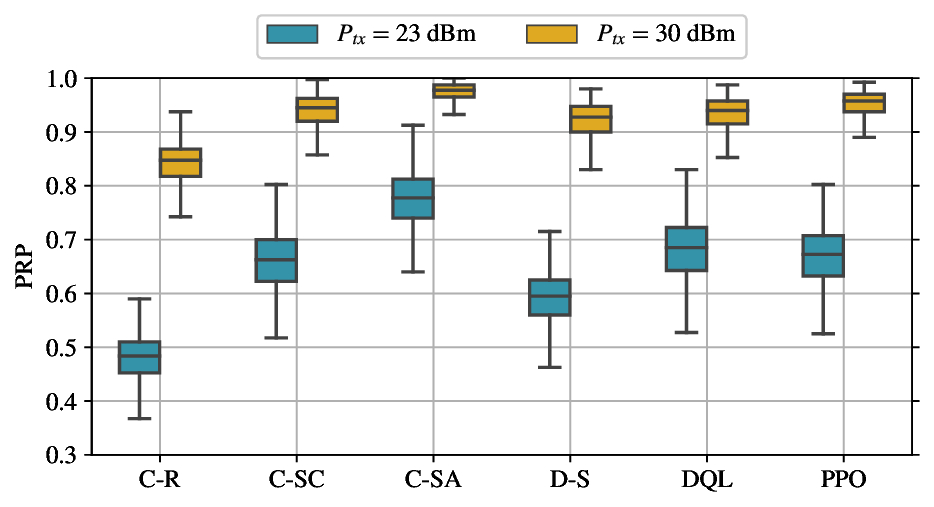}
        \caption{PRP distribution.}
        \label{fig:power_reception_bar}
    \end{subfigure}
    \caption{Communication metrics as a function of $P_{tx}$.}
    \label{fig:power_communication_metric}
    \vspace{-0.2cm}
\end{figure}

In Fig.~\ref{fig:power_communication_metric}, we represent the distribution of the end-to-end delay and \gls{prp} as a function of the transmission power.
Using $P_{tx}=23$ dBm leads to poor communication performance: the median \gls{prp} is below $0.70$ for all the strategies except C-SA, which keeps a value over $0.75$. 
The reduction in transmission power substantially affects end-to-end delay, increasing the interquartile range of the delay distribution, which reflects a higher jitter for the communication process. 
On the other hand, for \gls{ppo} the median delay is equal to $32$ ms in both conditions, denoting how the RAN-AI entity can address the system \glspl{kpi} even in critical scenarios.

\subsection{Impact of the State Space}

From the results presented in the first part of this section, we conclude that PRATA can well optimize PQoS via the RAN-AI entity, especially when the network conditions, in terms of number of users and transmission power, become more critical. 
In the following, we consider a scenario with $N_u=5$ and $P_{{tx}}=30$~dBm, and analyze the performance of the RAN-AI (trained according to either the \gls{dql} or \gls{ppo} algorithms) as a function of the state space.
Specifically, we consider the following options:
\begin{itemize}
    \item \gls{dql}$_{\text{app}}$ uses all the measurements generated at the application layer of the target vehicle, so the state space consists of $S=5$ variables;
    \item \gls{dql}$_{\text{phy}}$ uses all the measurements generated at the \gls{phy} and application layers of the target vehicle, so the state space consists of $S=8$ variables;
    \item \gls{dql}$_{\text{full}}$ and \gls{ppo}$_{\text{full}}$ use all the measurements generated at the \gls{phy}, \gls{pdcp}, \gls{rlc}, and application layers of the target vehicle, so the state space consists of $S=18$ variables. This is the configuration we considered in our previous results in Secs.~\ref{sub:size} and~\ref{sub:power}. In general, we expect \gls{dql}$_{\text{full}}$ and \gls{ppo}$_{\text{full}}$ to achieve a better perception of the environment given the large state space, and make more accurate decisions, despite the increase of the \gls{nnet}~size.
\end{itemize}
As anticipated in Sec.~\ref{sec:ran-ai}, we can obtain a more comprehensive description of the environment by including network contextual information within the state space. 
This is fundamental for ensuring a more robust training phase, and avoiding that the stochastic properties of \glspl{mdp} change during the simulation.
Hence, we consider two additional state configurations:
\begin{itemize}
    \item \gls{dql}$_{\text{app}}^{\text{net}}$ uses all the measurements at the application layer of the target vehicle, and the average measurements at the application layer of the other vehicles in the network, so the state space consists of $S=10$ variables.
    \item \gls{dql}$_{\text{phy}}^{\text{net}}$ uses all the measurements at the \gls{phy} and application layers of the target vehicle, and the average measurements at the \gls{phy} and application layers of the other vehicles in the network, so the state space consists of $S=16$ variables.
\end{itemize}
However, \gls{dql}$_{\text{app}}^{\text{net}}$ and \gls{dql}$_{\text{phy}}^{\text{net}}$ involve two critical points: the number of input neurons of the \gls{nnet} architecture increases, and the state includes variables that are out of the agent's control.
These aspects may increase the training complexity since the agent has to learn how to deal with a higher number of features with a more unpredictable evolution.

\begin{figure}[t!]
    \centering
    \includegraphics[width=.89\linewidth]{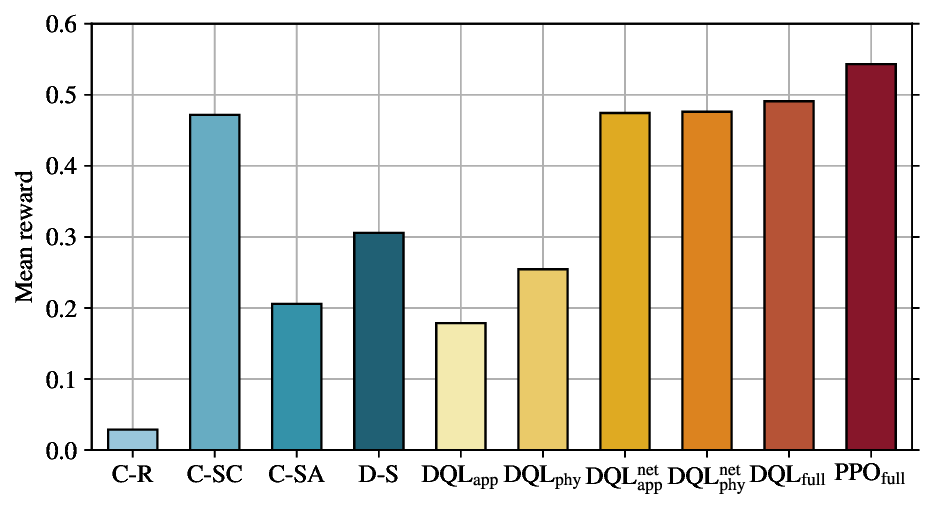}
    \caption{Mean reward as a function of $\mathcal{S}$.}
    \label{fig:state_reward_bar}
    \vspace{-0.2cm}
\end{figure}

In Fig.~\ref{fig:state_reward_bar} we report the mean reward for each state configuration, vs. the constant and heuristic benchmarks. 
As expected, \gls{dql}$_{\text{app}}$ has the worst results since state measurements from the application layer alone are insufficient to represent the complexity of the teleoperated network.
\gls{dql}$_{\text{phy}}$ slightly improves the performance by exploiting the metrics of the other layers of the communication stack.
A further performance enhancement is provided by \gls{dql}$_{\text{full}}$ while, as observed in the previous subsections, \gls{ppo}$_{\text{full}}$ proves to be the best approach in the target scenario.

Extending the state space using contextual information from other vehicles strongly improves the total reward: \gls{dql}$_{\text{app}}^{\text{net}}$ achieves a higher performance in terms of reward than \gls{dql}$_{\text{app}}$ (0.47 vs 0.18), and the same occurs for \gls{dql}$_{\text{phy}}^{\text{net}}$ against \gls{dql}$_{\text{phy}}$ (0.48 vs 0.25).
As discussed in Sec.~\ref{sec:ran-ai}, contextual information is fundamental to discern situations in which the status of the target vehicle is unchanged, but the other vehicles are under different conditions. 
This is not possible in the case of \gls{dql}$_{\text{app}}$ and \gls{dql}$_{\text{phy}}$ since the agent looks only at local variables, whose evolution changes as the training phase goes on.
In other words, \gls{dql}$_{\text{app}}$ and \gls{dql}$_{\text{phy}}$ exploit a much more noisy state space, making it more challenging for the RAN-AI entity to find the optimal policy.

\begin{figure}[t!]
    \centering
    \includegraphics[width=.89\linewidth]{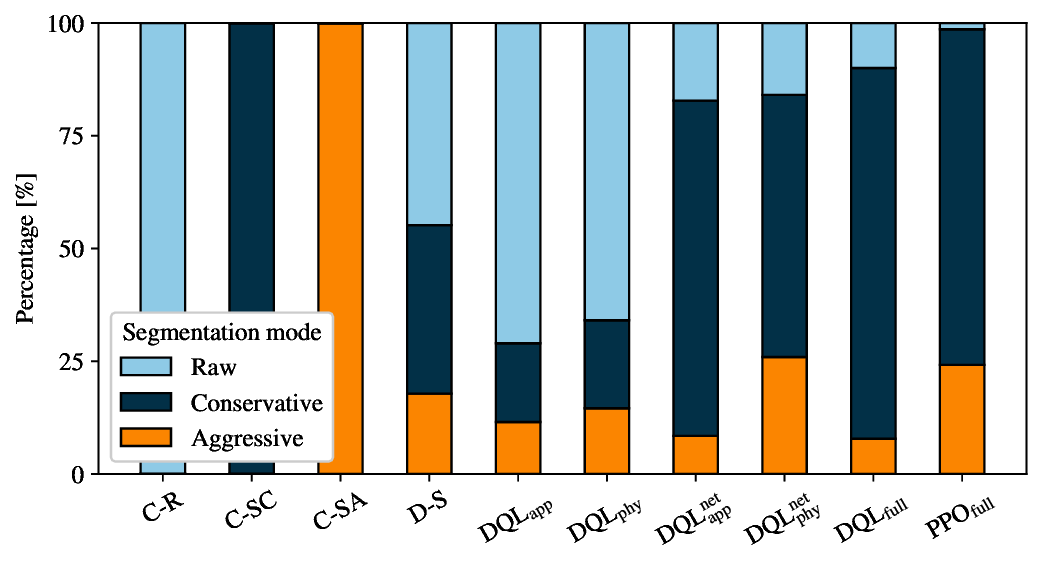}
    \caption{Action distribution as a function of $\mathcal{S}$.}
    \label{fig:state_qoe_hist}
    \vspace{-0.2cm}
\end{figure}

In Fig.~\ref{fig:state_qoe_hist} we report the distribution of the segmentation modes for different state space configurations.
We observe that \gls{dql}$_{\text{app}}$ mostly transmits raw data, and uses an aggressive segmentation approach for about $10\%$ of the time, similarly to \gls{dql}$_{\text{phy}}$.
In fact, both approaches are outperformed by \gls{ds}, which still scarcely uses the aggressive segmentation mode, but almost doubles the probability of adopting the conservative segmentation mode.
A marked increase in the probability of SC occurs in the case of \gls{dql}$_{\text{app}}^{\text{net}}$ and \gls{dql}$_{\text{phy}}^{\text{net}}$, with the latter adopting aggressive segmentation more frequently (more than $20$\% of the time). 
\gls{dql}$_{\text{full}}$ exploits SC more frequently than both \gls{dql}$_{\text{app}}^{\text{net}}$ and \gls{dql}$_{\text{phy}}^{\text{net}}$, which is consistent with the results in Fig.~\ref{fig:state_reward_bar}, while \gls{ppo}$_{\text{full}}$ generally selects both SC and SA segmentation, and hardly transmits raw data.

\begin{figure}[t!]
    \centering
    \begin{subfigure}{.89\linewidth}
        \centering
        \includegraphics[width=.99\linewidth]{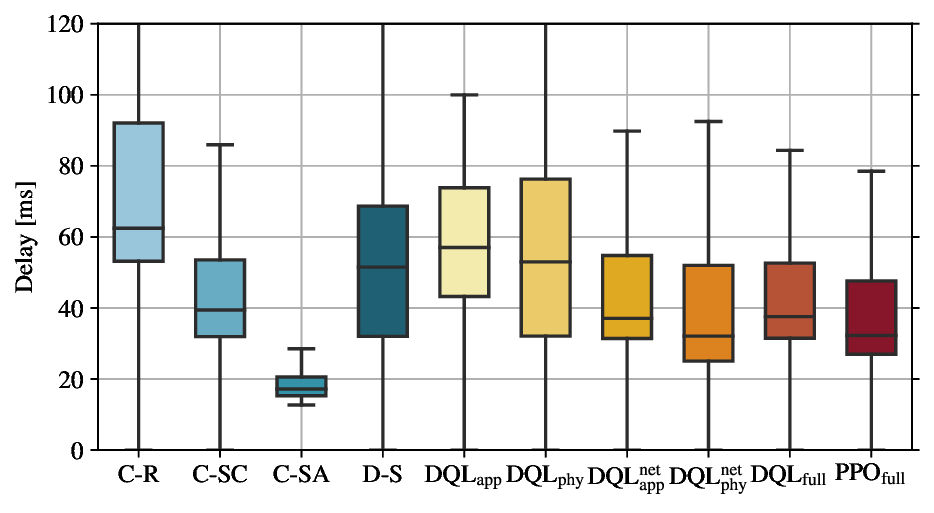}
        \caption{Packet delay distribution.}
        \label{fig:state_delay_box}
    \end{subfigure}\\
    \begin{subfigure}{.89\linewidth}
        \centering
        \includegraphics[width=.99\linewidth]{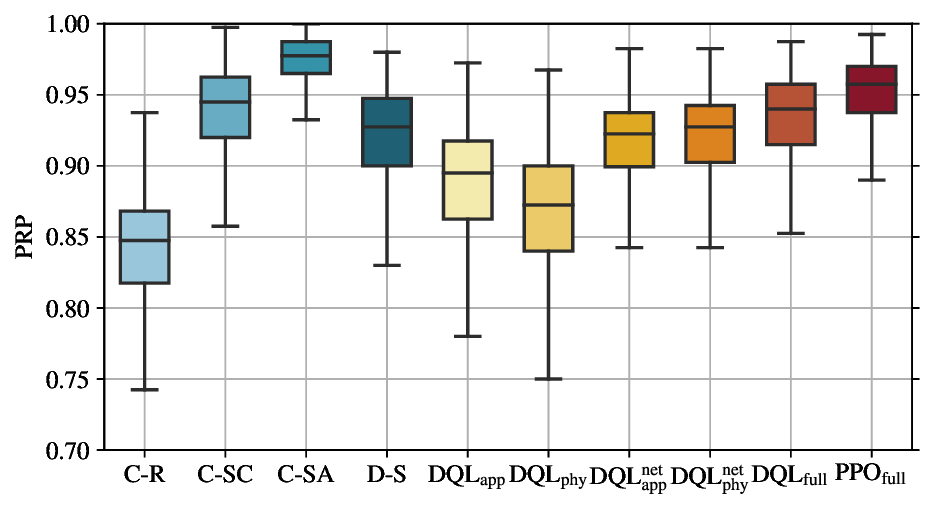}
        \caption{PRP distribution.}
        \label{fig:state_reception_box}
    \end{subfigure}
    \caption{Communication metrics as a function of $\mathcal{S}$.}
    \label{fig:state_communication_metric}
    \vspace{-0.2cm}
\end{figure}

Finally, Fig.~\ref{fig:state_delay_box} outlines the distributions of the end-to-end delay and \gls{prp} for the different state space configurations.
We observe that \gls{dql}$_{\text{app}}^{\text{net}}$, \gls{dql}$_{\text{phy}}^{\text{net}}$, and \gls{dql}$_{\text{full}}$ all ensure a delay lower than $40$ ms for more than $50$\% of the time.
Instead, \gls{dql}$_{\text{app}}$ and \gls{dql}$_{\text{phy}}$ do not address the delay requirements, leading to a median delay greater than $\delta_M=50$ ms. 
In particular, \gls{dql}$_{\text{phy}}$ comes with a higher percentile in the delay distribution than \gls{dql}$_{\text{app}}$ and the other learning approaches, which denotes a more severe jitter. 

This is consistent with the results in Fig.~\ref{fig:state_reception_box}, where the $75$-th percentile of the \gls{prp} distribution of \gls{dql}$_{\text{app}}^{\text{net}}$, \gls{dql}$_{\text{phy}}^{\text{net}}$, and \gls{dql}$_{\text{full}}$ are all above $0.90$, contrary to \gls{dql}$_{\text{app}}^{\text{net}}$ and \gls{dql}$_{\text{phy}}^{\text{net}}$.
We conclude that \gls{dql}$_{\text{full}}$ represents the most accurate state space configuration in the context of this work, as it maximizes the performance from both a \gls{qos} and \gls{qoe} perspective.
In particular, it achieves a 2.7$\times$ higher reward than \gls{dql}$_{\text{app}}$ (where the RAN-AI entity only uses measurements at the application layer), and a 1.9$\times$ higher reward than \gls{dql}$_{\text{phy}}$ (where the state space also includes \gls{phy} information).
Interestingly, \gls{dql}$_{\text{full}}$ also outperforms \gls{dql}$^{\text{net}}_{\text{app}}$ and \gls{dql}$^{\text{net}}_{\text{phy}}$, although the latter two have a more comprehensive perception of the network.

On the other hand, we observe that \gls{dql}$_{\text{full}}$, as well as \gls{ppo}$_{\text{full}}$, require aggregating measurements from multiple communication domains, at different layers of the protocol stack.
To this aim, it is necessary to define new application interfaces, which is not straightforward in the automotive domain.
Hence, there may be practical scenarios where a configuration with a smaller state space, such as \gls{dql}$_{\text{phy}}^{\text{net}}$, is a more suitable choice despite the lower theoretical performance. 
In addition, increasing the training time may reduce the performance gap between \gls{dql}$_{\text{full}}$ and \gls{dql}$_{\text{phy}}^{\text{net}}$, making the latter even more convenient.
At the same time, \gls{dql}$_{\text{phy}}^{\text{net}}$ presents the strong disadvantage of requiring the RAN-AI to aggregate data from different vehicles to compute the system state.
In a real scenario, this approach involves additional data transmissions to/from the gNB, which may increase the communication overhead and neutralize the benefits of using the RAN-AI.

\section{Conclusions}
\label{sec:conclusions}

In this paper we presented PRATA, a novel simulation framework for designing, evaluating, and dimensioning PQoS strategies for teleoperated driving applications.
We used PRATA to define an ad hoc intelligent architecture, named RAN-AI, that determines the optimal segmentation mode to apply to point cloud data transmitted from teleoperated vehicles to remote drivers.
Our simulation results showed that the RAN-AI entity efficiently optimizes the trade-off between \gls{qos} and \gls{qoe}, and almost doubled the system performance with respect to non-data-driven solutions, considering both static and adaptive segmentation modes.

We tested PRATA according to multiple network configurations and \gls{rl} algorithms, and observed that \gls{ppo} is more suitable to manage scenarios with multiple users than value-based \gls{rl} methods such as DQL, while the final performance depends on the communication measurements acquired by the RAN-AI.
In this regard, extending the state space involves critical challenges, such as the definition of new interfaces and a more significant communication overhead, which may not be convenient in this scenario.

As part of our future research activities, we will extend PRATA to incorporate more advanced functionalities, including multi-cell scenarios, as well as other countermeasures besides tuning the segmentation modes at the application. 
Finally, we are interested in extending our learning framework by considering new feature compression techniques and training architectures, e.g., based on \gls{fl}, and evaluating their benefits over a centralized system.

\bibliographystyle{IEEEtran}
\bibliography{./bibl}

\begin{thebibliography}{10}
\providecommand{\url}[1]{#1}
\csname url@samestyle\endcsname
\providecommand{\newblock}{\relax}
\providecommand{\bibinfo}[2]{#2}
\providecommand{\BIBentrySTDinterwordspacing}{\spaceskip=0pt\relax}
\providecommand{\BIBentryALTinterwordstretchfactor}{4}
\providecommand{\BIBentryALTinterwordspacing}{\spaceskip=\fontdimen2\font plus
\BIBentryALTinterwordstretchfactor\fontdimen3\font minus
  \fontdimen4\font\relax}
\providecommand{\BIBforeignlanguage}[2]{{%
\expandafter\ifx\csname l@#1\endcsname\relax
\typeout{** WARNING: IEEEtran.bst: No hyphenation pattern has been}%
\typeout{** loaded for the language `#1'. Using the pattern for}%
\typeout{** the default language instead.}%
\else
\language=\csname l@#1\endcsname
\fi
#2}}
\providecommand{\BIBdecl}{\relax}
\BIBdecl

\bibitem{giordani2020toward}
M.~Giordani, M.~Polese, M.~Mezzavilla, S.~Rangan, and M.~Zorzi, ``{Toward 6G
  Networks: Use Cases and Technologies},'' \emph{IEEE Communications Magazine},
  vol.~58, no.~3, pp. 55--61, March 2020.

\bibitem{letaief2019roadmap}
K.~B. Letaief, W.~Chen, Y.~Shi, J.~Zhang, and Y.-J.~A. Zhang, ``{The Roadmap to
  6G: AI Empowered Wireless Networks},'' \emph{IEEE Communications Magazine},
  vol.~57, no.~8, pp. 84--90, August 2019.

\bibitem{tong2019artificial}
W.~Tong, A.~Hussain, W.~X. Bo, and S.~Maharjan, ``{Artificial Intelligence for
  Vehicle-To-Everything: A Survey},'' \emph{IEEE Access}, vol.~7, pp.
  10\,823--10\,843, January 2019.

\bibitem{giordani2019investigating}
M.~Giordani, A.~Zanella, T.~Higuchi, O.~Altintas, and M.~Zorzi,
  ``{Investigating Value of Information in Future Vehicular Communications},''
  in \emph{IEEE Connected and Automated Vehicles Symposium (CAVS)}, 2019.

\bibitem{rossi2021role}
V.~Rossi, P.~Testolina, M.~Giordani, and M.~Zorzi, ``{On the Role of Sensor
  Fusion for Object Detection in Future Vehicular Networks},'' in \emph{Joint
  European Conference on Networks and Communications (EuCNC) \& 6G Summit},
  2021.

\bibitem{nardo2022point}
F.~Nardo, D.~Peressoni, P.~Testolina, M.~Giordani, and A.~Zanella, ``{Point
  Cloud Compression for Autonomous Driving: A Performance Comparison},'' in
  \emph{IEEE Wireless Communications and Networking Conference (WCNC)}, 2022.

\bibitem{zheng2015heterogeneous}
K.~Zheng, Q.~Zheng, P.~Chatzimisios, W.~Xiang, and Y.~Zhou, ``Heterogeneous
  vehicular networking: A survey on architecture, challenges, and solutions,''
  \emph{IEEE Communications Surveys \& Tutorials}, vol.~17, no.~4, pp.
  2377--2396, June 2015.

\bibitem{gyawali2020challenges}
S.~Gyawali, S.~Xu, Y.~Qian, and R.~Q. Hu, ``{Challenges and Solutions for
  Cellular Based V2X Communications},'' \emph{IEEE Communications Surveys \&
  Tutorials}, vol.~23, no.~1, pp. 222--255, October 2020.

\bibitem{kousaridas2021qos}
A.~Kousaridas, R.~P. Manjunath, J.~Perdomo, C.~Zhou, E.~Zielinski, S.~Schmitz,
  and A.~Pfadler, ``{QoS Prediction for 5G Connected and Automated Driving},''
  \emph{IEEE Communications Magazine}, vol.~59, no.~9, pp. 58--64, September
  2021.

\bibitem{boban2021predictive}
M.~Boban, M.~Giordani, and M.~Zorzi, ``{Predictive Quality of Service: The Next
  Frontier for Fully Autonomous Systems},'' \emph{IEEE Network}, vol.~35,
  no.~6, pp. 104--110, November 2021.

\bibitem{kaelbling1996reinforcement}
L.~P. Kaelbling, M.~L. Littman, and A.~W. Moore, ``{Reinforcement Learning: A
  Survey},'' \emph{{Journal of Artificial Intelligence Research}}, vol.~4, pp.
  237--285, May 1996.

\bibitem{sutton2018reinforcement}
R.~S. Sutton and A.~G. Barto, \emph{{Reinforcement Learning: An
  Introduction}}.\hskip 1em plus 0.5em minus 0.4em\relax MIT press, 2018.

\bibitem{xiong2019deep}
Z.~Xiong, Y.~Zhang, D.~Niyato, R.~Deng, P.~Wang, and L.-C. Wang, ``{Deep
  Reinforcement Learning for Mobile 5G and Beyond: Fundamentals, Applications,
  and Challenges},'' \emph{IEEE Vehicular Technology Magazine}, vol.~14, no.~2,
  pp. 44--52, June 2019.

\bibitem{osinski2020simulation}
B.~Osinski, A.~Jakubowski, P.~Zikecina, P.~Milos, C.~Galias, S.~Homoceanu, and
  H.~Michalewski, ``{Simulation-Based Reinforcement Learning for Real-World
  Autonomous Driving},'' in \emph{{IEEE International Conference on Robotics
  and Automation (ICRA)}}, 2020.

\bibitem{kumar2020conservative}
A.~Kumar, A.~Zhou, G.~Tucker, and S.~Levine, ``{Conservative Q-Learning for
  Offline Reinforcement Learning},'' in \emph{{Neural Information Processing
  Systems (NeurIPS)}}, 2020.

\bibitem{henderson2008network}
T.~R. Henderson, M.~Lacage, G.~F. Riley, C.~Dowell, and J.~Kopena, ``{Network
  Simulations with the ns-3 Simulator},'' in \emph{{ACM Special Interest Group
  on Data Communications (SIGCOMM)}}, 2008.

\bibitem{mezzavilla2018end}
M.~Mezzavilla, M.~Zhang, M.~Polese, R.~Ford, S.~Dutta, S.~Rangan, and M.~Zorzi,
  ``{End-to-End Simulation of 5G mmWave Networks},'' \emph{IEEE Communications
  Surveys \& Tutorials}, vol.~20, no.~3, pp. 2237--2263, April 2018.

\bibitem{zhao2022elite}
L.~Zhao, Z.~Bi, A.~Hawbani, K.~Yu, Y.~Zhang, and M.~Guizani, ``{ELITE: An
  Intelligent Digital Twin-based Hierarchical Routing Scheme for Softwarized
  Vehicular Networks},'' \emph{IEEE Transactions on Mobile Computing}, no.~9,
  pp. 5231--5247, May 2022.

\bibitem{mason2022rlpqos}
F.~Mason, M.~Drago, T.~Zugno, M.~Giordani, M.~Boban, and M.~Zorzi, ``{A
  Reinforcement Learning Framework for PQoS in a Teleoperated Driving
  Scenario},'' \emph{IEEE Wireless Communications and Networking Conference
  (WCNC) Workshops}, 2022.

\bibitem{drago2022wns3}
M.~Drago, T.~Zugno, F.~Mason, M.~Giordani, M.~Boban, and M.~Zorzi,
  ``{Artificial Intelligence in Vehicular Wireless Networks: A Case Study Using
  ns-3},'' in \emph{{ACM Workshop on ns-3 (WNS3)}}, 2022, p. 112–119.

\bibitem{khan2023ai}
N.~A. Khan and S.~Schmid, ``{AI-RAN} in {6G} networks state-of-the-art and
  challenges,'' \emph{IEEE Open Journal of the Communications Society}, vol.~5,
  pp. 294--311, December 2023.

\bibitem{9124820}
S.~K. Singh, R.~Singh, and B.~Kumbhani, ``The evolution of radio access network
  towards open-ran: Challenges and opportunities,'' in \emph{IEEE Wireless
  Communications and Networking Conference (WCNC) Workshops}, 2020, pp. 1--6.

\bibitem{10537699}
P.~Chauhan, N.~Sharma, and A.~Kumar, ``Artificial intelligence based vehicular
  networks: A step towards smart transportation system,'' in \emph{IEEE
  International Conference on Image Information Processing (ICIIP)}, 2023.

\bibitem{rizzo2023towards}
G.~Rizzo, E.~Liotou, Y.~Maret, J.-F. Wagen, T.~Zugno, M.~Wu, and A.~Kliks,
  ``Towards {AI}-native vehicular communications,'' in \emph{IEEE Vehicular
  Technology Conference (VTC)}, 2023.

\bibitem{9860971}
M.~Boban, C.~Jiao, and M.~Gharba, ``{Measurement-based Evaluation of Uplink
  Throughput Prediction},'' in \emph{{IEEE Vehicular Technology Conference
  (VTC)}}, 2022.

\bibitem{jomrich2018cellular}
F.~Jomrich, A.~Herzberger, T.~Meuser, B.~Richerzhagen, R.~Steinmetz, and
  C.~Wille, ``Cellular bandwidth prediction for highly automated driving,'' in
  \emph{{IEEE International Conference on Vehicle Technology and Intelligent
  Transport Systems (VEHITS)}}, 2018, pp. 121--132.

\bibitem{9129382}
L.~Torres-Figueroa, H.~F. Schepker, and J.~Jiru, ``{QoS Evaluation and
  Prediction for C-V2X Communication in Commercially-Deployed LTE and Mobile
  Edge Networks},'' in \emph{IEEE Vehicular Technology Conference (VTC)}, 2020.

\bibitem{9129097}
A.~Pfadler, G.~Jornod, A.~E. Assaad, and P.~Jung, ``{Predictive Quality of
  Service: Adaptation of Platoon Inter-Vehicle Distance to Packet
  Inter-Reception Time},'' in \emph{IEEE Vehicular Technology Conference
  (VTC)}, 2020.

\bibitem{10012940}
J.~Perdomo, M.~Gutierrez-Estevez, A.~Kousaridas, C.~Zhou, and J.~F. Monserrat,
  ``{QoS Prediction-based Radio Resource Management},'' in \emph{IEEE Vehicular
  Technology Conference (VTC)}, 2022.

\bibitem{10200750}
R.~Hernangómez~et al., ``Berlin v2x: A machine learning dataset from multiple
  vehicles and radio access technologies,'' in \emph{IEEE Vehicular Technology
  Conference (VTC)}, 2023, pp. 1--5.

\bibitem{10268872}
M.~Skocaj, N.~Di~Cicco, T.~Zugno, M.~Boban, J.~Blumenstein, A.~Prokes,
  T.~Mikulasek, J.~Vychodil, K.~Mikhaylov, M.~Tornatore, and V.~Degli-Esposti,
  ``Vehicle-to-everything (v2x) datasets for machine learning-based predictive
  quality of service,'' \emph{IEEE Communications Magazine}, vol.~61, no.~9,
  pp. 106--112, 2023.

\bibitem{9604941}
S.~Barmpounakis, L.~Magoula, N.~Koursioumpas, R.~Khalili, J.~M. Perdomo, and
  R.~P. Manjunath, ``{LSTM-based QoS prediction for 5G-enabled Connected and
  Automated Mobility applications},'' in \emph{IEEE 5G World Forum (5GWF)},
  2021, pp. 436--440.

\bibitem{9685405}
J.~Perdomo, A.~Kousaridas, C.~Zhou, and J.~F. Monserrat, ``{Deep Learning-based
  QoS Prediction with Innate Knowledge of the Radio Access Network},'' in
  \emph{IEEE Global Communications Conference (GLOBECOM)}, 2021.

\bibitem{9773810}
H.~Schippers, C.~Schüler, B.~Sliwa, and C.~Wietfeld, ``{System Modeling and
  Performance Evaluation of Predictive QoS for Future Tele-Operated Driving},''
  in \emph{{IEEE International Systems Conference (SysCon)}}, 2022.

\bibitem{https://doi.org/10.48550/arxiv.2302.11966}
A.~Palaios, C.~L. Vielhaus, D.~F. Külzer, C.~Watermann, R.~Hernangomez,
  S.~Partani, P.~Geuer, A.~Krause, R.~Sattiraju, M.~Kasparick, G.~Fettweis,
  F.~H.~P. Fitzek, H.~D. Schotten, and S.~Stanczak, ``{Machine Learning for QoS
  Prediction in Vehicular Communication: Challenges and Solution Approaches},''
  \emph{IEEE Access}, vol.~11, pp. 92\,459--92\,477, August 2023.

\bibitem{gemv2}
\BIBentryALTinterwordspacing
M.~Boban. (2014) {GEMV2: Geometry Based Efficient Propagation Model for V2V
  Communication}. [Online]. Available: \url{http://vehicle2x.net/}
\BIBentrySTDinterwordspacing

\bibitem{SUMO2012}
D.~Krajzewicz, J.~Erdmann, M.~Behrisch, and L.~Bieker, ``{Recent Development
  and Applications of SUMO - Simulation of Urban MObility},''
  \emph{International Journal On Advances in Systems and Measurements}, vol.~5,
  no.~3, pp. 128--138, December 2012.

\bibitem{bragato2023towards}
F.~Bragato, T.~Lotta, G.~Ventura, M.~Drago, F.~Mason, M.~Giordani, and
  M.~Zorzi, ``{Towards Decentralized Predictive Quality of Service in
  Next-Generation Vehicular Networks},'' \emph{IEEE Information Theory and
  Applications (ITA) Workshop}, 2023.

\bibitem{bragato2024federated}
F.~Bragato, M.~Giordani, and M.~Zorzi, ``{Federated Reinforcement Learning to
  Optimize Teleoperated Driving Networks},'' \emph{IEEE Global Communications
  Conference (GLOBECOM)}, 2024.

\bibitem{boban2014geometry}
M.~Boban, J.~Barros, and O.~K. Tonguz, ``{Geometry-Based Vehicle-to-Vehicle
  Channel Modeling for Large-Scale Simulation},'' \emph{IEEE Transactions on
  Vehicular Technology}, vol.~63, no.~9, November 2014.

\bibitem{piro2011lte}
G.~Piro, N.~Baldo, and M.~Miozzo, ``{An LTE Module for the ns-3 Network
  Simulator},'' in \emph{EAI International Conference on Simulation Tools and
  Techniques (ICST)}, 2011.

\bibitem{geiger2012are}
A.~Geiger, P.~Lenz, and R.~Urtasun, ``{Are We Ready for Autonomous Driving? The
  KITTI Vision Benchmark Suite},'' in \emph{IEEE Conference on Computer Vision
  and Pattern Recognition (CVPR)}, 2012.

\bibitem{varischio2021hybrid}
A.~Varischio, F.~Mandruzzato, M.~Bullo, M.~Giordani, P.~Testolina, and
  M.~Zorzi, ``{Hybrid Point Cloud Semantic Compression for Automotive Sensors:
  A Performance Evaluation},'' \emph{IEEE International Conference on
  Communications (ICC)}, 2021.

\bibitem{rnet}
A.~Milioto, I.~Vizzo, J.~Behley, and C.~Stachniss, ``{RangeNet++: Fast and
  Accurate LiDAR Semantic Segmentation},'' in \emph{IEEE/RSJ International
  Conference on Intelligent Robots and Systems (IROS)}, 2019.

\bibitem{lecci21bursty}
M.~Lecci, A.~Zanella, and M.~Zorzi, ``{An ns-3 Implementation of a Bursty
  Traffic Framework for Virtual Reality Sources},'' in \emph{{ACM Workshop on
  ns-3 (WNS3)}}, 2021.

\bibitem{perkis2020qualinet}
\BIBentryALTinterwordspacing
A.~Perkis, C.~Timmerer, S.~Barakovi{\'c}, J.~B. Husi{\'c}, S.~Bech, S.~Bosse,
  J.~Botev, K.~Brunnstr{\"o}m, L.~Cruz, K.~De~Moor \emph{et~al.}, ``{QUALINET
  white paper on definitions of immersive media experience (IMEx)},''
  \emph{arXiv}, 2020. [Online]. Available:
  \url{https://arxiv.org/abs/2007.07032}
\BIBentrySTDinterwordspacing

\bibitem{3GPP_22186}
{3GPP}, ``{Service requirements for enhanced V2X scenarios (Release 15)},''
  \emph{TS 22.186}, September 2018.

\bibitem{brunnstrom2013qualinet}
K.~Brunnstr{\"o}m, S.~A. Beker, K.~De~Moor, A.~Dooms, S.~Egger, M.-N. Garcia,
  T.~Hossfeld, S.~Jumisko-Pyykk{\"o}, C.~Keimel, M.-C. Larabi \emph{et~al.},
  ``Qualinet white paper on definitions of quality of experience,'' 2013.

\bibitem{van2016deep}
H.~Van~Hasselt, A.~Guez, and D.~Silver, ``{Deep Reinforcement Learning with
  Double Q-Learning},'' in \emph{{AAAI Conference on Artificial Intelligence}},
  2016.

\bibitem{zakharenkov2021deep}
A.~Zakharenkov and I.~Makarov, ``{Deep reinforcement learning with dqn vs. ppo
  in vizdoom},'' in \emph{IEEE International Symposium on Computational
  Intelligence and Informatics (CINTI)}, 2021.

\bibitem{pmlr-v48-mniha16}
V.~Mnih, A.~P. Badia, M.~Mirza, A.~Graves, T.~Lillicrap, T.~Harley, D.~Silver,
  and K.~Kavukcuoglu, ``{Asynchronous Methods for Deep Reinforcement
  Learning},'' in \emph{PMLR International Conference on Machine Learning
  (ICML)}, 2016.

\bibitem{mason2024multi}
F.~Mason, F.~Chiariotti, A.~Zanella, and P.~Popovski, ``Multi-agent
  reinforcement learning for coordinating communication and control,''
  \emph{IEEE Transactions on Cognitive Communication and Networking}, vol.~10,
  no.~4, pp. 1566--1581, August 2024.

\bibitem{chang1995cumulative}
T.~Chang and F.~Gan, ``A cumulative sum control chart for monitoring process
  variance,'' \emph{Journal of Quality Technology}, vol.~27, no.~2, pp.
  109--119, 1995.

\bibitem{yu2022surprising}
C.~Yu, A.~Velu, E.~Vinitsky, J.~Gao, Y.~Wang, A.~Bayen, and Y.~Wu, ``The
  surprising effectiveness of ppo in cooperative multi-agent games,''
  \emph{Advances in Neural Information Processing Systems (NeurlPS)}, 2022.

\end{thebibliography}

\end{document}